\documentclass{aa}

\usepackage{graphicx}
\usepackage{txfonts}
\begin{document}

   \title{The symbiotic binary St 2-22: Orbital and stellar parameters and jet evolution following its 2019 outburst}


   \author{Ga{\l}an C.\inst{1}
          \and Miko{\l}ajewska J.\inst{1}
          \and I{\l}kiewicz K.\inst{2}
          \and Monard B.\inst{3}
          \and \.Zywica S.\,T.\inst{4}
          \and Zamanov R.\,K.\inst{5}
          }

   \institute{Nicolaus Copernicus Astronomical Center, Polish Academy of Sciences, Bartycka 18, 00-716 Warsaw, Poland \\ \email{cgalan@camk.edu.pl} \label{inst1}
         \and Centre for Extragalactic Astronomy, Department of Physics, University of Durham, South Road, Durham DH1 3LE, UK \label{inst2}
         \and Kleinkaroo Observatory, Calitzdorp, Western Cape, South Africa \label{inst3}
         \and Institute of Astronomy, Faculty of Physics, Astronomy and Informatics, Nicolaus Copernicus University in Toru\'n, Grudzi\c{a}dzka 5, Toru\'n 87-100, Poland \label{inst4}
         \and Institute of Astronomy and National Astronomical Observatory, Bulgarian Academy of Sciences, Tsarigradsko Shose 72, 1784 Sofia, Bulgaria \label{inst5}
             }

   \date{Received 3 September 2021 / Accepted 22 October 2021}

 
  \abstract
   {St\,2-22 is a relatively poorly studied S-type symbiotic system that
belongs to a small group of jet-producing systems as a result of
disc accretion onto a white dwarf fed by its red giant companion.}
   {The goal of this paper is to analyse the nature and derive the basic
parameters of St\,2-22, and to follow the jet evolution.}
   {Photometric monitoring for over 16 yrs and high-quality spectroscopic
data enabled us to shed new light on its nature.  The high-resolution SALT
spectra and $V\,I_{\rm C}$ photometry obtained during and between the last
two outbursts have been used to search for periodic changes, to derive
spectroscopic orbits of both system components, and to study the outburst
and jet evolution.}
   {We present the orbital and stellar parameters of the system components. 
The orbital period is P$_{\rm orb} = 918$\,$\pm6^{\rm d}$.  The double-line
spectroscopic orbits indicate the mass ratio $q = M_{\rm g} M_{\rm h}^{-1} =
3.50$\,$\pm0.53$, and the components masses $M_{\rm g} \sin^3{i} \sim 2.35
$\,M$_{\sun}$ and $M_{\rm h} \sin^3{i} \sim 0.67 $\,M$_{\sun}$.  The orbit
shows significant eccentricity, $e = 0.16$\,$\pm0.07$.  The orbital
inclination is close to 70\degr.  During outbursts, accelerating and
decelerating jets are observed with changes in their radial velocity
component in a range from $\sim$\,1500 up to nearly $1800$\,km\,s$^{-1}$. 
St\,2-22 turned out to be a classical symbiotic system very similar to the
precursor of the group -- Z\,And.}
   {}


   \maketitle
%

\section{Introduction} \label{Sec-Int}

St\,2-22 = PN\,Sa\,3-22 ($\alpha_{2000}$:
$13^{\rm{h}}$\,$14^{\rm{m}}$\,$30.3^{\rm{s}}$, $\delta_{2000}$:
$-58$\degr\,$51'$\,$49.6"$) is a poorly studied classical symbiotic binary,
composed of an $\sim$\,M4-type red giant (RG) and a hot accreting white dwarf
(WD).  It is one of a handful of symbiotic systems (SySt) known to produce
high-velocity ($\sim$\,a few $\times10^3$\,km\,s$^{-1}$) collimated jets
during their outbursts \citep{Tom2017}.  However, so far only one outburst
has been recorded with only two spectra taken during it, and many issues
have yet to be addressed.

Initially, the object was classified as a planetary nebula \citep{San1976},
but the object was reclassified as SySt by \citet{All1984} who identified
the Raman scattered \ion{O}{vi} 6825\AA\, line in its spectrum. 
\citet{VWin1993} reported brightness variations at the end of the 1980s and
the beginning of the 1990s, however without details about their nature.  A
search for linear polarisation by \citet{Gar2003} gave a negative result. 
\citet{Zam2008} included St\,2-22 in their studies of rotational velocities
of the RGs in SySt.  \citet{MuSm1999} estimated an M4.5 spectral type for
the RG.  \citet{Mik1997} estimated the distance $\sim$\,5\,kpc as well as the
temperature T$_h \sim$\,54-100$\times$10$^3$\,K and luminosity
L$_h \sim$\,600\,L$_\sun$ of the WD.  Collimated, bipolar jets with an average
velocity of $\sim$\,1700\,km\,s$^{-1}$ were discovered in spectra collected
during an unnoticed outburst in 2005 \citep{Tom2017}.  The nature of the
outburst was similar to those observed in classical SySt.  The chemical
composition and physical parameters of the RG, $[$\ion{O}{iii}$]$ emission
line ratios, and infrared colours of St\,2-22 are consistent with an S-type
SySt \citep{Tom2017}.

At the beginning of 2019, another eruption began that is still ongoing. 
Here we discuss the characteristics of the last outburst phenomenon and
evolution of jets.  We also provide the orbital and physical parameters of
St\,2-22 based on spectra collected around the last two outbursts.

\begin{table}[t!]
\fontsize{7}{8.2}\selectfont
 \caption{Heliocentric radial velocities of the red giant and the
cF-absorption lines with their 1\,$\sigma$ errors, and the maximum $IP$
observed.}
\label{T_Sp_Obs}      
\centering          
\begin{tabular}{@{}c@{\hskip 5mm}c|@{\hskip 5mm}c@{\hskip 5mm}c@{\hskip 5mm}c@{\hskip 5mm}c@{}}
\hline\hline       
\vspace{-2mm}
\\
 Date [UT]  & HJD  & Phase\tablefootmark{$\star$}  & $IP_{\rm{max}}$ & RV$_{\rm{cool}}$  & RV$_{\rm{cF}}$  \\
 yyyy-mm-dd & $-2450000$ &                         & [eV]            & [km\,s$^{-1}$]    & [km\,s$^{-1}$]  \\
\hline
\vspace{-2mm}
\\
 2005-02-01 & 3402.8 & 0.272  &  55 & $37.98$\,$\pm0.32$  & --                \\
 2005-05-16 & 3506.6 & 0.385  &  99 & $36.46$\,$\pm0.37$  & --                \\
 2017-04-17 & 7861.3 & 0.129  & 114 & $36.24$\,$\pm0.69$  & --                \\
 2017-05-10 & 7884.5 & 0.154  & 114 & $39.28$\,$\pm1.00$  & --                \\
 2017-06-12 & 7917.4 & 0.190  & 114 & $40.01$\,$\pm0.64$  & --                \\
 2017-07-06 & 7941.3 & 0.216  & 114 & $37.82$\,$\pm0.67$  & --                \\
 2017-07-30 & 7965.3 & 0.242  & 114 & $39.55$\,$\pm0.69$  & --                \\
 2018-02-05 & 8154.5 & 0.448  & 114 & $33.32$\,$\pm0.76$  & --                \\
 2018-03-04 & 8182.4 & 0.479  & 114 & $30.21$\,$\pm1.25$  & --                \\
 2019-02-15 & 8530.5 & 0.858  &  55 & $25.91$\,$\pm0.68$  & $55.6$\,$\pm1.3$  \\
 2019-03-30 & 8572.6 & 0.904  &  25 & $26.75$\,$\pm0.49$  & $56.8$\,$\pm1.3$  \\
 2019-05-01 & 8605.5 & 0.940  &  25 & $29.01$\,$\pm0.33$  & $49.4$\,$\pm1.9$  \\
 2019-06-08 & 8643.4 & 0.981  &  25 & $31.26$\,$\pm0.44$  & $30.7$\,$\pm1.5$  \\
 2020-01-28 & 8876.5 & 0.235  &  55 & $39.44$\,$\pm0.49$  & $11.2$\,$\pm0.7$  \\
 2020-05-10 & 8980.5 & 0.348  &  55 & $36.29$\,$\pm0.50$  & $15.2$\,$\pm1.0$  \\
 2020-07-01 & 9032.3 & 0.404  &  55 & $35.40$\,$\pm0.43$  & $16.1$\,$\pm1.1$  \\
 2021-02-04 & 9249.5 & 0.641  &  55 & $29.54$\,$\pm0.64$  & $47.8$\,$\pm2.8$  \\
 2021-03-26 & 9300.4 & 0.697  &  55 & $26.22$\,$\pm0.43$  & $55.5$\,$\pm2.3$  \\
 2021-05-02 & 9337.3 & 0.737  &  55 & $24.17$\,$\pm0.37$  & $50.3$\,$\pm0.8$  \\
 2021-07-25 & 9421.3 & 0.828  &  55 & $25.02$\,$\pm0.47$  & --                \\
\hline
\end{tabular}
\tablefoot{\tablefoottext{$\star$}{Phase according to the ephemeris: $HJD_{\rm{T_0}} = 2458661 + 918 \times E$ (Table\,\ref{T_Orb_sol}).}
          }
\end{table}

 \begin{figure}[!h]
   \centering
   \includegraphics[height=0.60\textwidth]{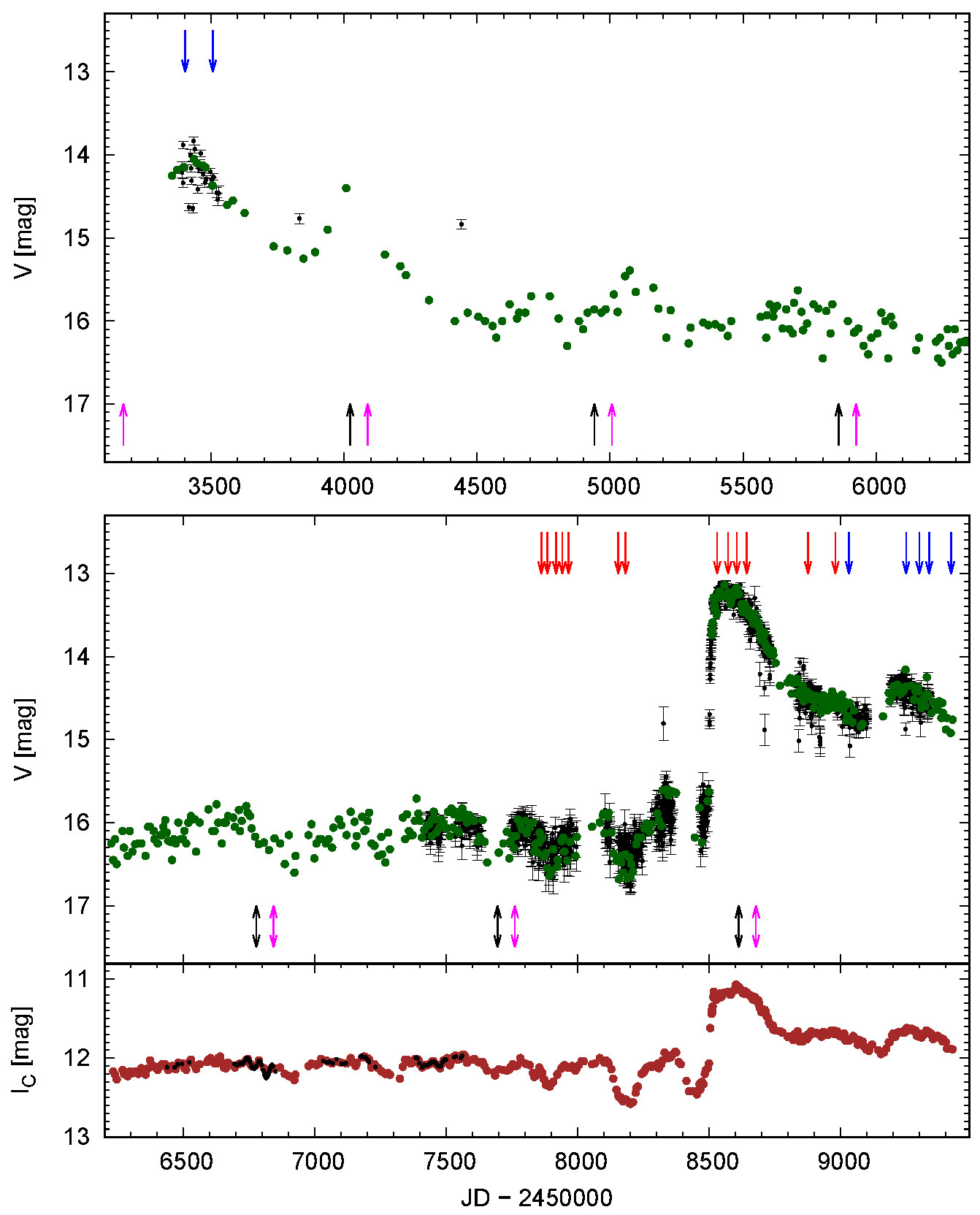}
      \caption{Light curve of St\,2-22 with the ongoing outburst.\ Green and
brown dots correspond to the $V$- and $I_{\rm C}$-band data, respectively,
acquired at the Kleinkaroo Observatory, whereas the {\sl ASAS} and {\sl
ASAS-SN} $V$- and $g$-band, as well as {\sl OGLE} $I$-band data, are plotted
as black points.  The zero-point of the OGLE photometry is shifted by
$-0.38$ mag.  Arrows above the light curves mark the {\sl HRS} and {\sl
FEROS} observations with ({\sl blue}) and without ({\sl red}) jet detection,
while those below mark the times of periastron passage ({\sl black}) and
inferior conjunction ({\sl magenta}) according to the eccentric orbit
solution (Table\,\ref{T_Orb_sol}).
      }
         \label{F_LCs}
   \end{figure}

\section{Observations} \label{Sec-Obs}

\subsection{Spectra} \label{Sec-Spe}

High-resolution (R $\sim$\,40000; range: 3920--8780\,\AA) spectra were
collected with the Southern African Large Telescope (SALT) High Resolution
Spectrograph (HRS) in medium--resolution mode under programmes
2018-2-SCI-021 and 2019-1-MLT-008 (PI: C.  Ga{\l}an).  These data are
complemented with publically available, HRS spectra acquired in
low--resolution mode (R $\sim$\,14000; useful range: 4000--8790\,\AA) under
programmes 2017-1-SCI-046 and 2017-2-SCI-044 (PI: T.  Tomov).  In this case,
three exposures were collected for each night which have been finally summed
into single spectra.  The HRS observations were reduced using the
{\sl{MIDAS}}--based
pipeline\footnote{\tiny{http://www.saao.ac.za/$\sim$akniazev/pub/HRS$\_$MIDAS/HRS$\_$pipeline.pdf}}
\citep{Kni2016, Kni2017}.  We also used four high-resolution (R $\sim$\,48000;
3710--9215\,\AA) {\sl FEROS} spectra obtained with the 2.2-m MPG/ESO
telescope at La Silla Observatory under the programme 074.D-0114.  The
journal of our spectroscopic observations is presented in
Table\,\ref{T_Sp_Obs}.

\subsection{Photometry} \label{Sec-Pho}

St\,2-22 was monitored for over 16 years starting in December\,2004
(JD\,2453353) with a 35\,cm Meade\,RCX400 telescope equipped with an SBIG
ST8-XME CCD camera and $V$ and $I_{\rm{C}}$ (beginning in November\,2012;
JD\,2456232) filters at the Kleinkaroo Observatory (South Africa).  Each
single data point is the result of several individual exposures, which were
calibrated (dark subtraction and flat-fielding) and stacked selectively. 
Magnitudes were derived from differential photometry to nearby reference
stars using the single image mode of AIP4 image processing software.  Our
photometry was supplemented by the data in $V$ and $g$ filters from the All
Sky Automated Survey for Supernovae \citep[{\sl ASAS-SN},][]{Sha2014,
Koc2017} and by {\sl OGLE} $I$-band measurements \citep[see][]{Tom2017}. 
Moreover, the $V$-band data collected by the {\sl ASAS} survey
\citep{Poj1997} are used for comparison.  The light curves are shown in
Fig.\,\ref{F_LCs} with arrows marking the times of our spectroscopic
observations.

\section{Results and discussion} \label{Sec-AnDis}

\subsection{Optical light curves}

The light curves of St\,2-22 show a systematic brightness decline (by
$\Delta V$\,$\sim$\,$2$ over $\sim$\,2000 days) followed by about 10 yrs of
a stable $V$\,$\sim$\,$16$ with moderate periodic-like fluctuations, and
with a major outburst around JD\,24558505 (January 21, 2019) that is still
ongoing.  This outburst began with a steep brightening by $\Delta
V$\,$\sim$\,$2^{\rm m}$ ($\Delta I$\,$\sim$\,0.7) in $\la$\,$10^{\rm d}$,
followed by a slower brightening over a month up to $V_{\rm
max}$\,$\approx$\,$13.2$ ($I_{\rm max}$\,$\approx$\,$11.1$).  The steady
decline began after a short (JD\,2458537--614) optical plateau, initially by
$\Delta V$\,$\sim$\,$1$ ($\Delta I$\,$\sim$\,$0.6$) within about 180 days,
and then, around JD 2458860 St\,2-22 has come to a standstill at
$V$\,$\sim$\,$14.5$ ($I$\,$\sim$\,$11.7$), with small amplitude fluctuations
on timescales of $\sim 300$ days.

 \begin{figure*}[t!]
   \centering
   \includegraphics[height=0.58\textwidth]{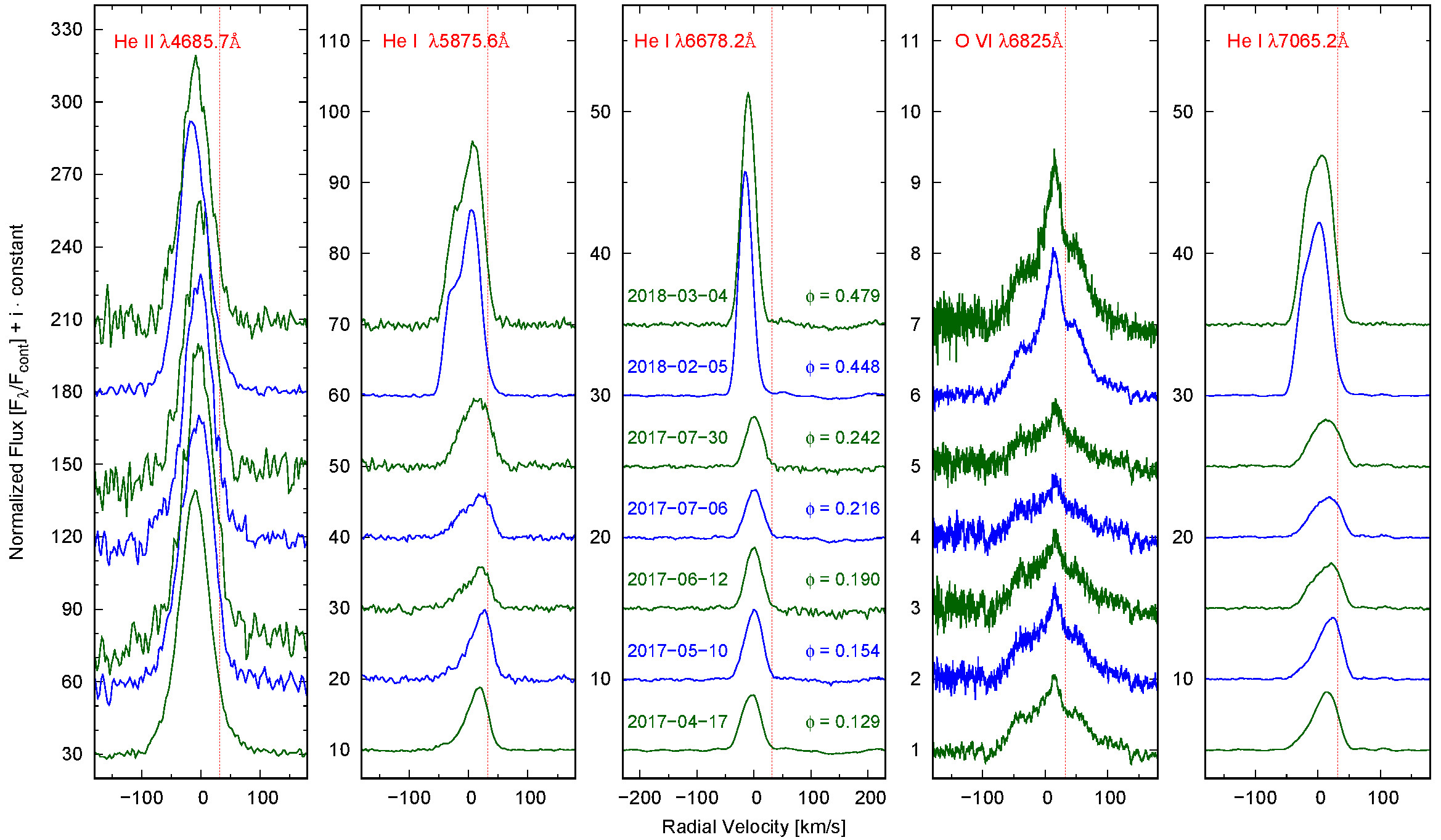}
      \caption{Evolution of the \ion{He}{ii} ($\lambda 4685.680$\,\AA) and
\ion{He}{i} ($\lambda 5875.640$\,\AA, $\lambda 6678.151$\,\AA, and $\lambda
7065.190$\,\AA) line profiles, and the Raman scattered \ion{O}{vi} ($\lambda
6825$\,\AA) line with the radial velocity scale corresponding to the parent
\ion{O}{vi} ($\lambda 1032$\,\AA) line in the HRS/SALT spectra
(Low-resolution mode: R $\sim$\,14000) collected during quiescence.  The red
dashed vertical lines mark the systemic velocity.}
         \label{F_He_II_I_q}
   \end{figure*}

While \citet{Tom2017} needed support from spectroscopy in addition to
photometry to prove the occurrence of the outburst, our light curve confirms
the reality of the 2005 outburst.  \citet{Tom2017} also reported the
outburst as relatively short, $\la$\,$200^{\rm d}$.  However, they did not
observed the outburst onset, and they assumed an incorrect quiescent
brightness of $V$\,$=$\,$15\fm32$.  In fact, St\,2-22 was not detected by
{\sl ASAS} before September 2004 which implies $V$\,$\ga$\,$15$ (i.e.  below
the detection limit of {\sl ASAS}) but then it remained invisible for
$\sim$\,$5$ months (September -- January) due to its position in the sky,
and it was detected at $V$\,$\sim$\,$14.2$ just when is became visible
again.  Hence, it is very likely that the 2005 outburst occurred during the
invisibility gap.

To search for any periodic changes, we analysed $V$-band light curve from
the quiescent phase (JD\,2454416--8283) using the discrete Fourier transform
method in the {\sl Period\,04} programme \citep{LeBe2005}.  We found two
periods, P$_{\rm 1}$\,$=$\,$891^{\rm d}$\,$\pm15$, which could be attributed
to the orbital cycle, and P$_{\rm 2}$\,$=$\,$277\fd8$\,$\pm1.7$, which is
most likely due to RG pulsation.  However, there is no evidence for a much
shorter period of $51^{\rm d}$\,$\pm7$ reported by \citet{Tom2017}.  The
analysis of the $I_{\rm C}$-band light curve confirmed the periods found in
the $V$-band data within $3$\,$\sigma$ error.  Phased light curves are
presented in Fig.\,\ref{F_ph_LCs}\,(Appendix).

\subsection{Spectral changes}

HRS spectra cover all the phases of recent activity -- the maximum, the
standstill following the early decline, and the substantial part of the
quiescent state (see Fig.\,\ref{F_LCs}).  Quiescent spectra of St\,2-22 are
characteristic of most SySt, with prominent emission lines from \ion{H}{i},
\ion{He}{i}, \ion{He}{ii}, [\ion{O}{iii}] as well as high ionisation
[\ion{Fe}{vii}] lines and the Raman scattered \ion{O}{vi} line
(Fig.\,\ref{F_He_II_I_q}), and a strong red continuum with deep TiO
absorption bands.  The highest ionisation potential ($IP _{\rm max}$)
observed in the spectrum is given in Table\,\ref{T_Sp_Obs} while the
behaviour of the main emission line profiles is shown in
Figs.\,\ref{F_HaHb_q}\,and\,\ref{F_He_II_I_q}, and the equivalent widths are
found in Table \ref{T_emission_lines_I}.  The strength of \ion{H}{i} and
\ion{He}{i} lines increases significantly in the spectra taken in 2018,
preceding the outburst by almost a year.  At the same time, the \ion{He}{ii}
lines remain unchanged.

\begin{table*}[t!]
\fontsize{7}{8.2}\selectfont
\caption{Journal of spectroscopic observations with information about time
(UT, HJD, phase) and exposure times.  }
\label{T_emission_lines_I}      
\centering          
\begin{tabular}{@{}c@{\hskip 5mm}c@{\hskip 5mm}c@{\hskip 4mm}c|@{\hskip 4mm}c@{\hskip 4mm}c@{\hskip 4mm}c@{\hskip 4mm}c@{\hskip 4mm}c@{\hskip 4mm}c@{\hskip 4mm}c@{\hskip 4mm}c@{\hskip 4mm}c@{\hskip 4mm}c@{}}
\hline\hline       
\vspace{-2mm}
\\
 Date [UT]\tablefootmark{$\blacklozenge$}  & HJD\tablefootmark{$\blacklozenge$}  & Phase\tablefootmark{$\star$}  & Exp. t.             & \ion{He}{ii}     & H$_\beta$     & [\ion{O}{iii}]   & [\ion{Ca}{vii}]  & \ion{He}{i}      & [\ion{Fe}{vii}] & H$_\alpha$    & \ion{He}{i}      & \ion{O}{vi}    & \ion{He}{i}      \\
                                           &                                     &                               &                     & $\lambda 4685.7$ &               & $\lambda 5006.9$ & $\lambda 5618.8$ & $\lambda 5875.6$ & $\lambda 6087$  &               & $\lambda 6678.2$ & $\lambda 6825$ & $\lambda 7065.2$ \\
 yyyy-mm-dd\,hh:mm:ss                      & $-2450000$                          &                               & $[$sec.$]$          & [\AA]            & [\AA]         & [\AA]            & [\AA]            & [\AA]            & [\AA]           & [\AA]         & [\AA]            & [\AA]          & [\AA]            \\
\hline
\vspace{-2mm}
\\
 2005-02-01\,06:16:51  & 3402.762  & 0.272  & 2$\times$1800  & 9    & 32  & 11   & 0 & 9    & 0    & 230 & 6        & 0  & 7  \\
 2005-05-16\,02:04:17  & 3506.591  & 0.385  & 2$\times$1800  & 14   & 30  & 12   & 0 & 9    & 0    & 261 & 9        & 0  & 9  \\
 2017-04-17\,19:48:06  & 7861.329  & 0.129  & 3$\times$1200  & 98   & 74  & 5    & 3 & 8    & 4    & 202 & 4        & 13 & 5  \\
 2017-05-10\,23:01:42  & 7884.464  & 0.154  & 3$\times$400   & 191: & 94: & 3    & 4 & 9    & 4    & 265 & 4        & 14 & 5  \\
 2017-06-12\,20:53:02  & 7917.373  & 0.190  & 3$\times$400   & nc   & nc  & 0    & 2 & 5    & 5    & 256 & 4        & 11 & 4  \\
 2017-07-06\,19:55:52  & 7941.333  & 0.216  & 3$\times$400   & nc   & nc  & nc   & 2 & 7    & 3    & 229 & 3        & 7  & 4  \\
 2017-07-30\,18:31:00  & 7965.272  & 0.242  & 3$\times$400   & nc   & nc  & nc   & 3 & 14   & 3    & 246 & 3        & 8  & 4  \\
 2018-02-05\,00:40:20  & 8154.528  & 0.448  & 3$\times$1200  & 111  & 150 & nc   & 4 & 26   & 4    & 549 & 10       & 19 & 16 \\
 2018-03-04\,22:28:27  & 8182.438  & 0.479  & 3$\times$600   & nc   & nc  & 0    & 5 & 33   & 5    & 587 & 12       & 24 & 16 \\
 2019-02-15\,23:18:25  & 8530.472  & 0.858  & 2340           & 0    & 8   & 1    & 0 & 2    & 0    & 107 & 1        & 0  & 1  \\
 2019-03-30\,02:00:56  & 8572.587  & 0.904  & 2200           & 0    & 9   & 0    & 0 & 0    & 0    & 113 & 0        & 0  & 0  \\
 2019-05-01\,23:58:17  & 8605.503  & 0.940  & 2570           & 0    & 7   & 0    & 0 & 0    & 0    & 109 & 0        & 0  & 0  \\
 2019-06-08\,21:06:40  & 8643.383  & 0.981  & 2950           & 0    & 13  & 0    & 0 & 3    & 0    & 175 & 2        & 0  & 2  \\
 2020-01-28\,00:36:18  & 8876.525  & 0.235  & 2750           & 15   & 26  & 8    & 0 & 11   & 0    & 218 & 8        & 0  & 9  \\
 2020-05-10\,22:47:55  & 8980.454  & 0.348  & 2750           & 14   & 32  & 10   & 0 & 11   & 0    & 228 & 8        & 0  & 10 \\
 2020-07-01\,19:39:40  & 9032.322  & 0.404  & 2750           & 13   & 39  & 10   & 0 & 12   & 0    & 243 & 9        & 0  & 10 \\
 2021-02-04\,00:35:29  & 9249.525  & 0.641  & 2900           & 10   & 32  & 7    & 0 & 7    & 0    & 189 & 7        & 0  & 7  \\
 2021-03-26\,21:21:13  & 9300.393  & 0.697  & 2900           & 14   & 53  & 10   & 0 & 11   & 0    & 268 & 9        & 0  & 9  \\
 2021-05-02\,18:17:17  & 9337.266  & 0.737  & 2600           & 10   & 43  & 9    & 0 & 9    & 0    & 245 & 7        & 0  & 9  \\
 2021-07-25\,18:15:44  & 9421.262  & 0.828  & 2600           & 19   & 54  & 13   & 0 & 15   & 0    & 332 & 12       & 0  & 14 \\
\hline                  
\end{tabular}
\tablefoot{The right-hand columns separated by a vertical line show the
measured equivalent widths (EWs) of emission lines.  The errors are of the
order of 1\,\AA.  The sign 'nc' means that EW was not measured, although the
line was visible because of a problem with estimating the continuum level
due to low S/N.  A colon designates the measurements that can be burdened
with large uncertainty and are difficult to estimate.
           \tablefoottext{$\blacklozenge$}{Time at mid of exposure.}\,
           \tablefoottext{$\star$}{Phase according to the ephemeris: $HJD_{\rm{T_0}} = 2458661 + 918 \times E$ (Table\,\ref{T_Orb_sol}).}
          }
\end{table*}

 \begin{figure}[hb!]
   \centering
\includegraphics[height=0.585\textwidth]{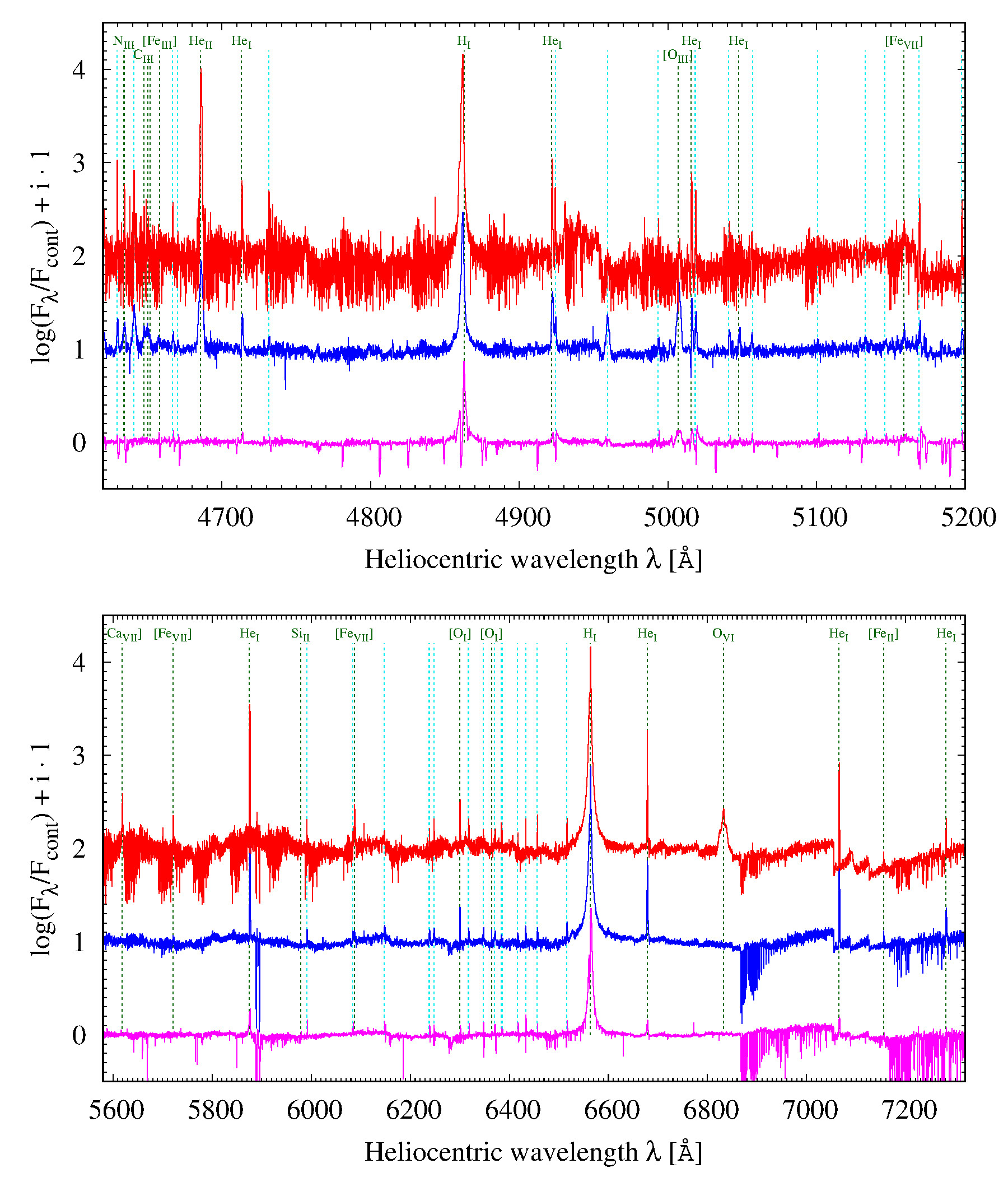}
      \caption{Parts of blue ({\sl top}), and red-arm ({\sl bottom}),
HRS/SALT spectra obtained close to the outburst maximum on February 15, 2019
(magenta), when the jets were first noticed on July 1, 2020 (blue), and
during quiescence on February 5, 2018 (red).  The laboratory wavelengths for
identified emission lines are shown with green dashed lines labelled with
the corresponding element.  The position of the most frequent lines from
singly ionised iron (\ion{Fe}{ii}) are shown with cyan dashed lines without
labels for clarity.}
         \label{F_emmision_lines}
   \end{figure}

The outburst amplitude and timescales, as well as spectral changes, resemble
those observed in other classical SySt (e.g.  Z\,And, CI\,Cyg, and AX\,Per),
the so-called Z\,And-type outbursts.  In all epochs, in the observed members
of the Balmer series, the broad emission is cut by a central absorption,
blueshifted by $\sim$30-40\,km\,s$^{-1}$ with respect to the systemic
velocity (Figs.~\ref{F_HaHb_q}\,and\,\ref{F_HaHb_o}).  The lines become stronger
during the outburst, and the absorption component seems to be broader,
especially in the spectrum closest to the outburst maximum (February 15,
2019).  All \ion{He}{i} and \ion{He}{ii} lines completely or nearly
disappear during the brightest phase of the outburst.  The red giant
absorption features (e.g.  TiO bands) weaken, and at times (e.g.  on
July\,1,\,2020) almost completely disappear (Fig.\,\ref{F_emmision_lines}). 
The spectra in the blue range are dominated by a hot continuum with
absorption lines mostly from \ion{Ti}{ii}, \ion{Fe}{ii}, and \ion{Cr}{ii}
which most closely resembles spectra of A--F supergiants.  This so-called
cF-shell absorption system is believed to be linked to the hot companion and
formed either in a thick accretion disc around the compact object
\citep{MiKe1992, Bra2005} or in the stream of accreted matter \citep[e.g.  in
AR\,Pav, RS\,Oph, \& V3890\,Sgr;][]{Sch2001, Qui2002, Bra2009, Mik2021}.

The spectra taken in 2020, after the significant drop in optical brightness,
reveal the reappearance of \ion{He}{ii} emission lines, while the blue
continuum and cF-shell absorption lines are still present, although
significantly weakened.  However, the most remarkable change occurred in
July 2020 when two satellite components with a velocity
$\sim$\,$\pm1650$\,km\,s$^{-1}$ appeared in H$\alpha$ wings, which we
attribute to a launch of bipolar jets.  The jets were detected for the first
time since their discovery during the 2005 outburst \citep{Tom2017}, and
they have been present in all subsequent spectra.

The {\sl FEROS} spectra obtained in 2005 resemble the HRS spectra taken
during the light curve standstill (2020 to 2021), and, in particular, the
spectrum taken on February 1, 2005 is almost identical with the HRS spectrum on
April 2, 2021 (almost exactly during a small secondary maximum).  This
suggests that both outbursts could have been similar and that the 2005
outburst had very likely started much earlier and reached a higher amplitude
than reported by \citet{Tom2017}.

\subsection{Spectroscopic orbits} \label{Sec-orb-sol}

All our spectra in the red region contain plenty of absorption lines from
the atmosphere of the cool, red giant.  We used the synthetic spectra from
the BT-NextGen grid of the theoretical spectra of
\citet{All2011}\footnote{http://svo2.cab.inta-csic.es/theory/newov2/index.php}{,
who used the solar abundances of \citet{Asp2009}, convolved with the
appropriate profile, to meet the resolution of the observed spectra.  The
synthetic spectrum of an M5\,III star ($T_{\rm {eff}} = 3400$\,K, $\log{g} =
0.5$, and $z=-0.25$) best fitted the observed quiescent spectra (see
Fig.\,\ref{F_Teff_class}).  To measure the radial velocities, our spectra
were cross-correlated with the synthetic spectrum.  To avoid the regions
with emission lines or polluted by the telluric features from absorption of
the Earth's atmosphere, the following masks were applied: 6110--6130,
6133--6147, 6597--6607, 6624--6656, 7069--7154, 7378--7513, 7899--8138, and
8670--8700\,\AA.  The resulting velocities are given in
Table\,\ref{T_Sp_Obs} and plotted in Fig.\,\ref{F_RVCs_2}.

Table\,\ref{T_cF_lines} lists  more than 100 cF-shell absorption system
lines, located in the range $\sim$\,$4200$--$5200$\,\AA, which were
indentified in the Feros and HRS spectra taken from 2019 to 2021 and used to
measure the radial velocities (Table\,\ref{T_Sp_Obs}).  The radial
velocities of these lines are in anti-phase with the red giant
(Fig.\,\ref{F_RVCs_2}), and we attribute them to the orbital motion of the WD.

Table\,\ref{T_Orb_sol} shows the orbital solutions.  The obtained orbital
period $P_{\rm {sp}}=918$\,$\pm6^{\rm d}$ is consistent within 3$\sigma$
errors with that from periodic analysis of the light curves.  Due to its
better precision, we adopt it as the final value.  The best solution gives a
significantly eccentric orbit (see Fig.\,\ref{F_RVCs_2}), although the
periastron longitude, $\omega$\,$\sim$\,270\degr, may indicate some
geometrical effect tied to the line of sight as in AR\,Pav \citep{Qui2002},
FN\,Sgr \citep{Bra2005}, and V3890\,Sgr \citep{Mik2021}.  However, in
St\,2-22 the same eccentric solution fits both radial velocity curves which
suggests that the orbit is indeed eccentric.  The semi-amplitudes of the
radial velocity curves of both components indicate the mass ratio
$q$\,$=$\,$3.50$\,$\pm0.53$, the masses $M_{\rm g}
\sin^3{i}=2.35$\,$\pm0.55$ M$_{\sun}$, and $M_{\rm WD} \sin^3{i}=
0.67$\,$\pm0.15$M$_{\sun}$, and the separation of the components
$a$\,$\sin{i}$\,$\sim$\,$2.67$\,AU.  The lack of eclipses sets the upper
limit for the orbit inclination, $i$\,$\la$\,$70\degr$, whereas the lower
limit, $i$\,$\ga$\,$52\degr$, results from the fact that the WD cannot
exceed the Chandrasekhar limit ($M_{\rm WD}$\,$\la$\,$1.4$M$_\sun$).  This
indicates a moderately massive WD, $M_{\rm
WD}$\,$\ga$\,$0.8$\,$\pm$\,$0.2$M$_\sun$.  The red giant mass is then
$M_{\rm g}$\,$\ga$\,$2.8$\,$\pm$\,$0.7$M$_\sun$.

 \begin{figure}[t!]
   \centering
\includegraphics[height=0.48\textwidth]{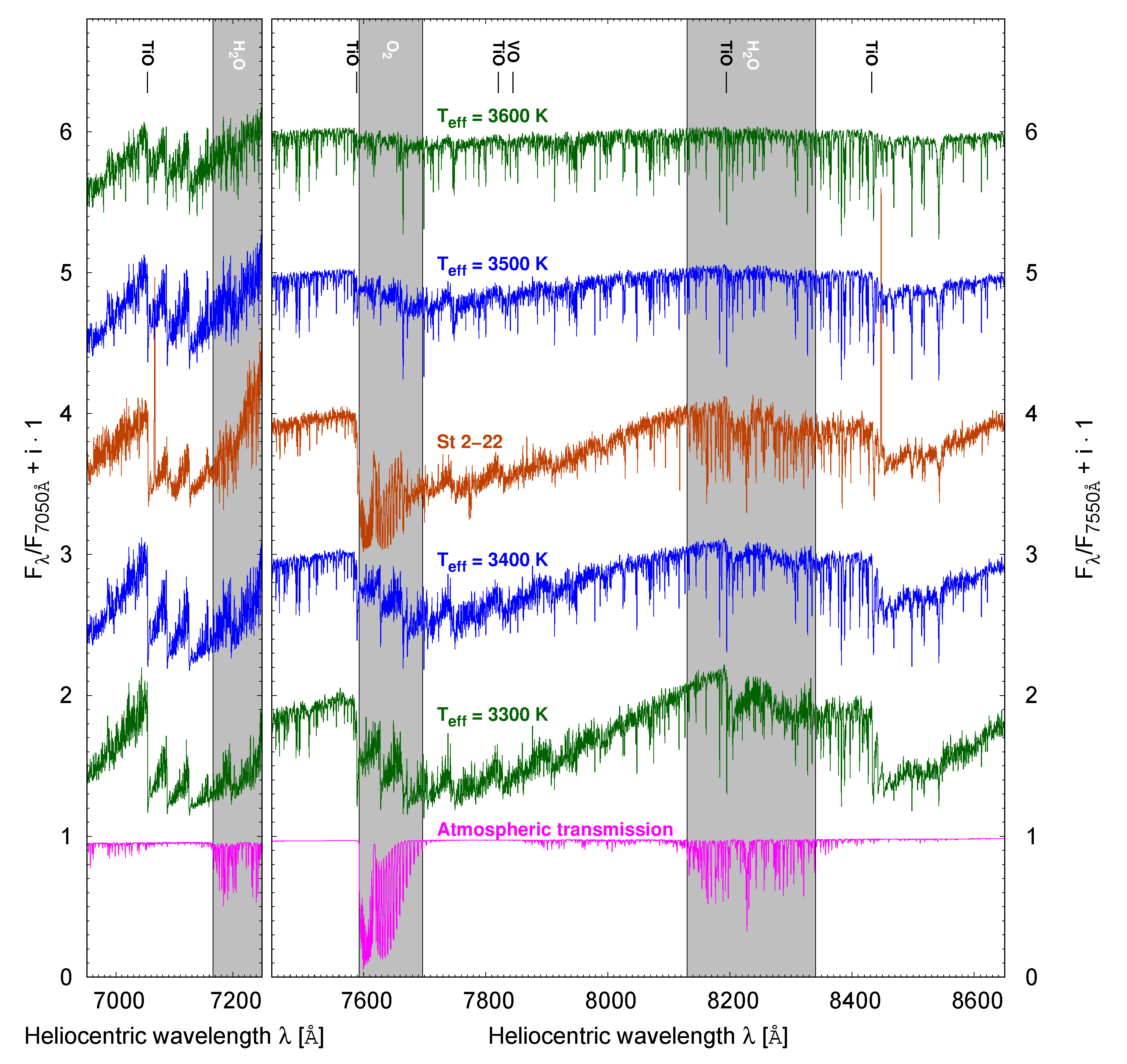}
      \caption{Average of three spectra of St\,2-22 obtained during
quiescence in June and July 2017, compared with the spectra from the grid
BT-NextGen of the theoretical spectra \citep[][$\log{g} = 0.5$\,dex and
$T_{\rm{eff}}$ in the range 3300--3600\,K]{All2011}.  The spectrum of
St\,2-22 most resembles the giant's spectrum with $T_{\rm{eff}} = 3400$\,K. 
The spectrum of the atmospheric transmission is shown at the bottom in
magenta; it was not subtracted from the observed spectra.  It was generated
with the use of the {\sl TAPAS}\protect\footnotemark\, service
\citep{Ber2014}, adopting conditions similar to those during our
observations in Sutherland.}
         \label{F_Teff_class}
 \end{figure}
 \footnotetext{\tiny{\url{http://cds-espri.ipsl.fr/tapas/}}}

\subsection{Red giant and distance} \label{Sec-dis-par}

Gaia\,DR3 gives a parallax of $\pi$\,$=$\,$0.085$\,$\pm0.023$\,mas
\citep{Gai2021} for St\,2-22, resulting in a value of a distance
$d$\,$=$\,$6.88^{+1.01}_{-0.97}$\,kpc \citep{Bai2021}.  The goodness-of-fit
statistic parameter for this parallax is {\sl gofAL}\,$\approx$\,2.05, which
means a good fit to the data, thus giving credibility to the measured distance. 
The {\sl 2MASS} colours \citep{Phi2007} and the mean quiescent
$V$\,$-$\,$I_{\rm c}$\,$\sim$\,$4.2$ (this paper) are consistent with
M5\,III ($T_{\rm eff}$\,$=$\,$3400$\,K; sec.\,\ref{Sec-orb-sol}) and the
reddening $E_{\rm B-V}$\,$=$\,$0.7$\,$\pm0.1$, which agrees with the total
Galactic extinction $E_{\rm B-V}$\,$<$\,$0.80$\,$\pm0.04$ \citep{ScFi2011}
and is consistent with $E_{\rm {B-V}}$\,$=$\,$0.81$\,$\pm0.10$
\citep{Zam2021}, as well as marginally consistent with $E_{\rm
{B-V}}$\,$\sim$\,$1$\,$\pm0.3$ estimated by \citet{Mik1997}.  We adopt the
distance $d$\,$=$\,$6.9$\,kpc and $E_{\rm B-V}$\,$=$\,$0.7$ for the rest of
the paper.

The reddening-corrected $K_0$\,$=$\,$7.97$\,$\pm0.05$ and $(J-K)_0$\,$=$
$1.20$\,$\pm0.09$ combined with the corresponding bolometric correction
$BC_{\rm K}$\,$=$\,$2.94$\,$\pm0.10$ \citep{BeWo1984} gives $M_{\rm
{bol}}$\,$=$\,$-3.28$\,$\pm0.45$, the luminosity $L_{\rm
g}$\,$=$\,$1631^{+835}_{-555}$\,L$_\sun$, and the radius $R_{\rm
g}$\,$=$\,$117$\,$\pm20$\,R$_{\sun}$.  The mass ratio, $q$\,$\sim$\,$3.5$,
$a$\,$\sin$\,$i$\,$\sim$\,$2.7$\,AU, and $\sin$\,$i$\,$\la$\,$70\degr$
(sec.\,\ref{Sec-orb-sol}), gives the Roche lobe radius $R_{\rm
RL}$\,$=$\,$0.5 a$\,$\sim$\,$306$\,R$_{\sun}$ \citep[][Eq.\,(4)]{Pac1971}
which indicates that the giant does not fill its tidal lobe even at the
periastron ($R_{\rm g}/R_{\rm RL,p} \sim 0.5$).

\citet{Hut1981} showed that in a binary with an eccentric orbit, the tidal
force acts to synchronise the donor's rotation with the orbital motion of
the companion, and the equilibrium (called pseudo synchronisation) is
reached for the rotational period, $P_{\rm rot}$, lower than $P_{\rm orb}$
by an amount depending on the eccentricity.  Using the orbital parameters
for St\,2-22 (Table~\ref{T_Orb_sol}), we estimated \rm{$P_{\rm rot}/P_{\rm
orb}$\,$=$\,$0.86$\,$\pm$0.09} \citep[][Eq.\,(42)]{Hut1981} which combined
with the RG rotational velocity
$v\sin{i}$\,$=$\,$9.8$\,$\pm1.5$\,km\,s$^{-1}$ \citep{Zam2008} gives the RG
radius $R_{\rm g}$\,$=$\,$153$\,$\pm40$\,R$_\sun$, in good agreement with
that derived above.

The luminosity and effective temperature ($\log T_{\rm eff}$\,$=$\,$3.53$,
$\log L_{\rm g}$\,$=$\,$3.21$\,$\pm0.20$) locate the giant on the HR diagram
between the evolutionary tracks of 1.6 and 2.5\,M$_{\sun}$
\citep[e.g.][]{Hur2000}, which more or less agree with our dynamical mass
estimate (sec.\,\ref{Sec-orb-sol}).  The systemic velocity
$\gamma$\,$=$\,$32.58$\,km\,s$^{-1}$ and proper motions $\mu_\alpha
\cos{\delta}$\,$=$\,$-5.713$\,$\pm0.018$\,mas\,yr$^{-1}$ and
$\mu_\delta$\,$=$\,$1.085$\,$\pm0.022$\,mas\,yr$^{-1}$ imply the following
Galactic velocities: $U$\,$=$\,$-136$, $V$\,$=$\,$-138$, and
$W$\,$=$\,$-16.3$\,km\,s$^{-1}$ which via confrontation with the Toomre
diagram \citep[see, e.g.][-- fig.\,1]{Fel2003} suggest that St\,2-22 may
belong to the extended thick-disc population, despite being placed
relatively close to the Galactic plane $z$\,$=$\,$0.46$\,$\pm0.08$\,kpc
($b$\,$=$\,$3.87$\degr).

 \begin{figure}[t!]
   \centering
   \includegraphics[height=0.45\textwidth]{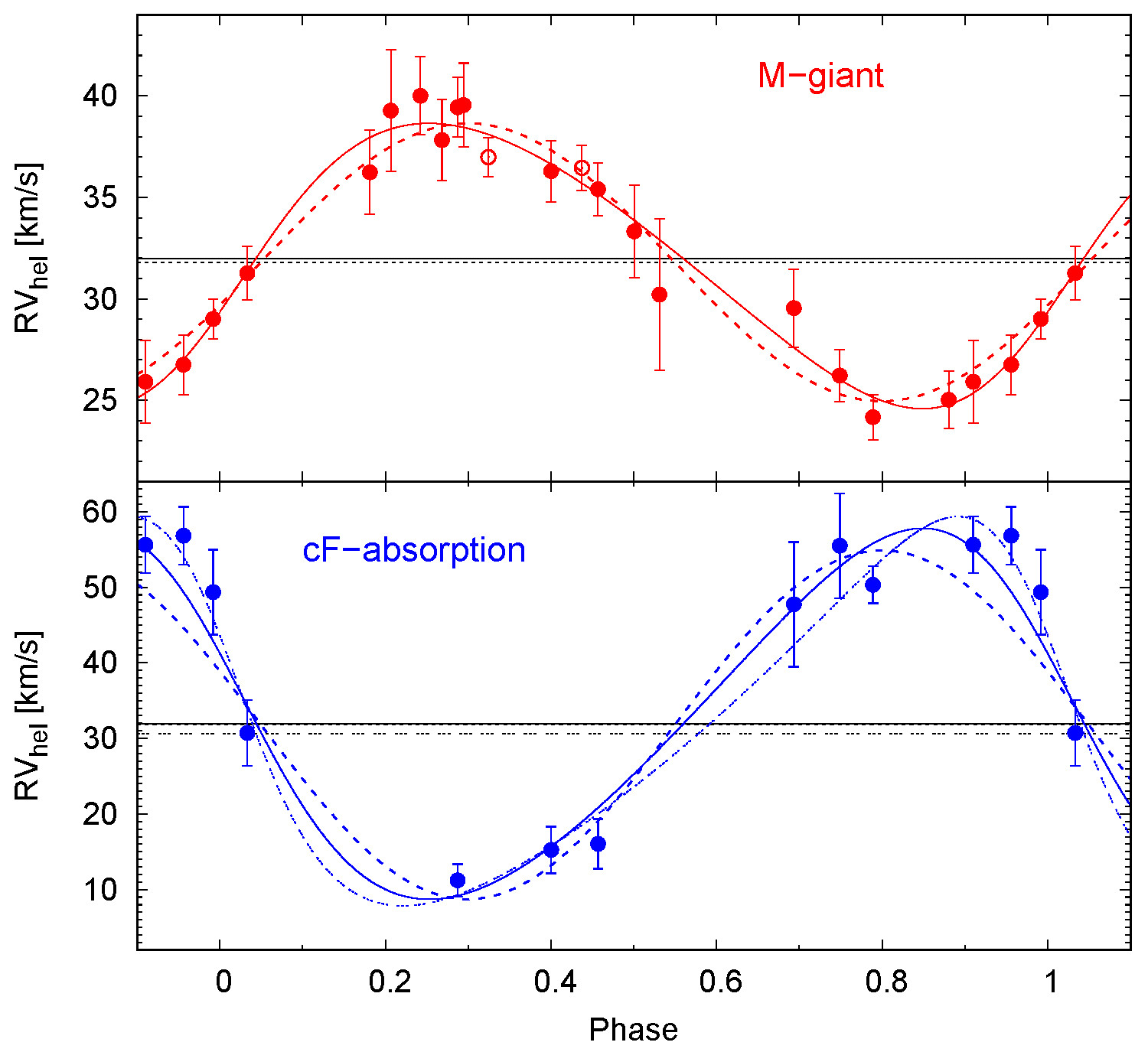}
      \caption{Radial velocity curves of the cool ({\sl {Top}}) and hot
               ({\sl {Bottom}}) components folded with the period P$_{\rm {sp}}
               = 918^{\rm d}$.  The values measured from SALT/HRS and FEROS data
               are shown with filled and open circles, respectively.  Lines
               show synthetic radial velocity curves: for the case of the
               eccentric orbit ({\sl solid}), circular orbit ({\sl dashed}),
               and from the solution for RV$_{\rm{cF}}$ only ({\sl
               dot dashed}).  The corresponding horizontal lines represent
               the systemic velocities in these three cases.}
         \label{F_RVCs_2}
   \end{figure}

\begin{table*}[t!]
\fontsize{7}{8.2}\selectfont
\caption{Orbital solutions for St\,2-22.}
\label{T_Orb_sol}      
\centering
\begin{tabular}{@{}l@{\hskip 5.5mm}c@{\hskip 5.5mm}c@{\hskip 5.5mm}c@{\hskip 5.5mm}c@{\hskip 5.5mm}c@{\hskip 5.5mm}c@{\hskip 5.5mm}c@{\hskip 5.5mm}c@{\hskip 5.5mm}c@{\hskip 5.5mm}c@{}}
\hline\hline       
\vspace{-2mm}
\\
P$_{\rm{orb}}$    & $V_\gamma$             & $K_{\rm g}$             & $K_{\rm h}$            & $q$                               & $e$               & $\omega$       & \,\,$T_0$\tablefootmark{a} & $T_{\rm{ic}}$\tablefootmark{b} & $f(m)$ & $a \sin{i}$ \\

$[$day$]$         & \small{[km\,s$^{-1}$]} &\small{[km\,s$^{-1}$]}   & \small{[km\,s$^{-1}$]} & [$M_{\rm{g}}$\,$M_{\rm{h}}^{-1}$] &                   & [deg]          & \multicolumn{2}{l}{[JD$ - 2450000$]}                       & [M$_{\sun}$]          & [AU]              \\
\hline
\vspace{-2mm}
\\
$918.0$\,$\pm6.0$ & $31.98$\,$\pm 0.30$    & $7.03$\,$\pm0.45$       & $24.6$\,$\pm1.9$       & $3.50$\,$\pm0.53$                 & $0.16$\,$\pm0.07$ & $251$\,$\pm22$ & $8613$\,$\pm52$                      & $8661$              & $0.0331$\,$\pm0.0066$ & $2.67$\,$\pm0.55$ \\
$918.0$\tablefootmark{c}& $30.66$\,$\pm 1.74$ &-- & $25.8$\,$\pm2.8$       & --                                & $0.29$\,$\pm0.09$ & $67$\,$\pm27$  & $8613$\,$\pm49$                      & --                  & --                    & --                \\
$922.1$\,$\pm6.4$ & $31.80$\,$\pm 0.34$    & $6.85$\,$\pm0.44$       & $23.1$\,$\pm2.4$       & $3.37$\,$\pm0.61$                 & 0                 &  --            & --                                   & $8658.8$\,$\pm10.4$ & $0.0307$\,$\pm0.0062$ & $2.54$\,$\pm0.60$ \\
\hline
\end{tabular}
\tablefoot{The errors are 1\,$\sigma$ values.
           \tablefoottext{a}{$T_0$ is a time of the periastron passage for the eccentric orbit.}\,
           \tablefoottext{b}{$T_{\rm {ic}}$ is a time of inferior conjunction.}\,
           \tablefoottext{c}{Additional solution for RV$_{\rm{cF}}$ only, adopting an orbital period from the eccentric solution when velocities of both components were used.}
          }
\end{table*}

\begin{table*}[h!]
\fontsize{7}{8.2}\selectfont
\caption{Radial velocities (RV) of the jets, FWHM values, and equivalent
widths (EW) measured in the HRS and FEROS spectra.}
\label{T_jet_RV}      
\centering          
\begin{tabular}{@{}c@{\hskip 5mm}c@{\hskip 5mm}c@{\hskip 5mm}c@{\hskip 5mm}c@{\hskip 5mm}c@{\hskip 5mm}c@{\hskip 5mm}c@{\hskip 5mm}c@{\hskip 5mm}c@{\hskip 5mm}c@{\hskip 5mm}c@{\hskip 5mm}c@{}}
\hline\hline       
\vspace{-2mm}
\\
  HJD               & Phase\tablefootmark{a}  & V$_{\rm{J}}^-$ & FWHM$^-$ & EW$^-_{\lambda}$ & V$_{\rm{J}}^+$ & FWHM$^+$ & EW$^+_{\lambda}$ & V$_{\rm{J}}$\tablefootmark{b} & VC$_{\rm{J}}$\tablefootmark{c} & EW$^-_{\lambda}/$EW$^+_{\lambda}$ & $V$ & $I_{\rm C}$ \\
 \small{$-2450000$} &                                       & \small{[km\,s$^{-1}$]} & \small{[km\,s$^{-1}$]} & \small{[\AA]} & \small{[km\,s$^{-1}$]} & \small{[km\,s$^{-1}$]} & \small{[\AA]} & \small{[km\,s$^{-1}$]} & \small{[km\,s$^{-1}$]} & & [mag] & [mag]  \\
\hline
\vspace{-2mm}
\\
 3402.76            & $0.272$  & $-1578$  & $380$  & $1.4$  & $1534$  & $338$  & $1.3$  & $1556$  & $-22$   & $1.1$  & $14.15$  & --      \\ 
 3506.59            & $0.386$  & --       & --     & --     & $1752$  & $332$  & $1.0$  & $1752$  & --      & --     & $14.37$  & --      \\ 
 9032.32            & $0.404$  & $-1642$  & $297$  & $1.1$  & $1688$  & $234$  & $0.7$  & $1665$  & $23  $  & $1.6$  & $14.76$  & $11.78$ \\ 
 9249.52            & $0.642$  & $-1464$  & $164$  & $0.8$  & $1558$  & $237$  & $0.5$  & $1511$  & $47  $  & $1.7$  & $14.36$  & $11.62$ \\ 
 9300.39            & $0.696$  & $-1604$  & $223$  & $1.0$  & $1789$  & $201$  & $0.5$  & $1697$  & $92.5$  & $1.8$  & $14.54$  & $11.73$ \\ 
 9337.27            & $0.736$  & $-1604$  & $339$  & $2.0$  & $1795$  & $324$  & $1.1$  & $1700$  & $95.5$  & $1.8$  & $14.48$  & $11.71$ \\ 
 9421.26            & $0.828$  & $-1588$  & $316$  & $1.5$  & $1692$  & $268$  & $0.9$  & $1640$  & $52$    & $1.7$  & $14.84$  & $11.88$ \\ 
\hline                  
\end{tabular}
\tablefoot{The 1\,$\sigma$ uncertainties are of the order of 10--20\,km\,s$^{-1}$ for RV and FWHM, and $0.1$--$0.2$ for EW.
           \tablefoottext{a}{Phase according to the ephemeris: $HJD_{\rm{T_0}} = 2458661 + 918 \times E$ (Table\,\ref{T_Orb_sol}).}\,
           \tablefoottext{b}{V$_{\rm{J}} = (|$V$_{\rm{J}}^-| + |$V$_{\rm{J}}^+|)/2$.}\,
           \tablefoottext{c}{VC$_{\rm{J}} = ($V$_{\rm{J}}^- + |$V$_{\rm{J}}^+)/2$.}
          }
\end{table*}

\subsection{Hot component activity and jet evolution}

The optical magnitudes recorded for St\,2-22 was $<$\,$V$\,$>_{\rm hot}
\approx 13.25$ during the optical plateau (JD\,2458537--614), which after
correcting for the RG contribution ($V_{\rm g}$\,$\approx$\,$16.2$)
corresponds to $V_{\rm hot}$\,$\approx$\,$13.3$.  Assuming that most of the
hot component continuum emission is shifted to the optical (a lower limit to
luminosity if not), and that during outburst $m_{\rm
bol}$\,$\approx$\,$V_{\rm hot}$, we estimate the reddening corrected $m_{\rm
bol}$\,$\approx$\,$11.1$, and the absolute bolometric magnitude $M_{\rm
bol}$\,$\approx$\,$ -3.1$, which corresponds to $L_{\rm
hot}$\,$\approx$\,$1380$\,L$_{\sun}$.

Similarly, we estimate the luminosity of the F-type component, $L_{\rm
hot}$\,$\sim$\,$300$--$400$\,L$_{\sun}$, during the standstill from 2020 to
2021 and in 2005, respectively.  In addition, to the F-type features, the
spectra display very strong \ion{H}{i}, \ion{He}{i}, and \ion{He}{ii}
emission lines.  Unfortunately, the SALT spectra are not flux-calibrated,
thus we cannot estimate the emission line fluxes and the corresponding
temperature and luminosity of the ionising source.  The {\sl {FEROS}}
spectra have relative flux-calibration, and we applied an absolute flux
scale to the observed $V$\,mag (Table~\ref{T_jet_RV}) such that convolution
of the spectrum with the Johnson $V$ filter agrees with the $V$\,mag.  The
emission line fluxes measured on the calibrated {\sl {FEROS}} spectra are
$F(H\beta)$\,$=$\,$7.7$ and
$7.4$\,$\times$\,$10^{-14}$\,erg\,s$^{-1}$\,cm$^{-2}$,
$F(\ion{He}{i}\,5876)$\,$=$\,$9.3$ and
$8.3$\,$\times$\,$10^{-14}$\,erg\,s$^{-1}$\,cm$^{-2}$, and
$F(\ion{He}{ii}\,4686)$\,$=1.2$ and
$2.0$\,$\times$\,$10^{-14}$\,erg\,s$^{-1}$\,cm$^{-2}$ on February\,1 and
May\,5,\,2005, respectively.  These values are significantly different from
those reported by \citet{Tom2017} (table 1): For example, the reddened
(using their $E_{\rm B-V}$\,$=$\,$1$) values of $F(H\beta)$\,$=$\,$1.0$ and
$2.8$\,$\times$\,$10^{-13}$\,erg\,s$^{-1}$\,cm$^{-2}$ from February\,1 and
May\,5,\,2005, respectively, are larger than ours, and, moreover, the fluxes
measured in May are $\sim$\,$3$ times larger than those in February, whereas
both the observed $V$ mag and the EWs (Tables\,\ref{T_emission_lines_I} and
\ref{T_jet_RV}) are practically identical.  Apparently,
there is something wrong with the absolute flux calibration of the
{\sl{FEROS}} spectra in \citet{Tom2017}.  Assuming a blackbody spectrum and
case B recombination, the fluxes of H$\beta$, \ion{He}{i}\,5876, and
\ion{He}{ii}\,4686 lines require $T_{\rm h}$\,$\sim$\,$100$\,kK (also
indicated by the presence of weak [\ion{Fe}{vii}] emission lines in the
best-exposed spectra), and luminosity (within a factor of 2), $L_{\rm
h}$\,$\sim$\,$200$\,L$_\sun$ \citep[see, e.g.][]{Mer2020}.  The estimated
luminosity is similar to the quiescent $L_{\rm
h}$\,$\approx$\,$220$\,L$_\sun$ resulting from the emission line fluxes
published by \citet{Mik1997} corrected for the reddening and distance
adopted in this study (Sec.\,\ref{Sec-dis-par}), although the line ratios
indicate a higher temperature, $T_{\rm h}$\,$\sim$\,$140$\,kK.  Based on the
lack of Raman scattered lines in this spectrum, \citet{Tom2017} suggested
that this spectrum was obtained during another missed outburst.  However,
the visual magnitude, $V$\,$\approx$\,$15.5$\,$\pm0.3$, that we derived from
this spectrum is more or less consistent with quiescent magnitudes
(Figs\,\ref{F_LCs}\,and\,\ref{F_ph_LCs}), and the lack of the Raman
\ion{O}{vi} and [\ion{Fe}{vii}] lines is probably due to insufficient
sensitivity to detect these relatively faint lines of this observation.

The simultaneous presence of the F-type component and the much hotter
ionising source points to a double-temperature structure of the active hot
component that can be associated with an accretion disc.  The F-type shell
spectrum would be then formed in the optically thick disc seen nearly
edge-on ($i \la 70$), while the hot radiation produced in a boundary layer
and/or in the innermost disc regions would photoionise low density material
above and below the disc and give rise to strong emission lines.  A similar
interpretation was proposed for other active SySt which simultaneously show
an A/F-type shell spectrum and high ionisation emission lines
\citep[e.g.][]{KeWe1984, Qui2002, Mik2021}.  The luminosities estimated for
both components agree within a factor of two which supports such an
interpretation.  Finally, the presence of an accretion disc is also
compatible with the jet production.

Figure\,\ref{F_Ha} shows the evolution of the jet components in the
H$\alpha$ line profile.  The jets are clearly visible as two satellite
emission components at velocity V$_{\rm J}$\,$\sim$\,$\pm1650$\,km\,s$^{-1}$
in the spectrum acquired in July 2020 (JD\,2459032), while they were absent
in that taken in May (JD\,2458980).  This indicates that the jets were
launched somewhere between May and July, shortly after St\,2-22 reached
a standstill in the optical light curve.  These additional components are
present in all subsequent spectra.

Table\,\ref{T_jet_RV} presents radial velocities (measured by fitting
Gaussian profiles) of both -- approaching (V$_{\rm{J}}^-$) and receding
(V$_{\rm{J}}^+$) -- jets together with the values of full widths at half
maximum (FWHM), equivalent width (EW), and other characteristics.  The
outflow velocities, FWHMs, and EWs vary with time although there are no
obvious trends.  The approaching component is always stronger than the
receding one with EW$^-_{\lambda}/$EW$^+_{\lambda}$\,$\sim$2 from 2020-2021. 
The slowest jets were observed on JD\,2459249, and their FWHM and EW are the
lowest there as well.  On the contrary, during the 2005 outburst, the
slowest jets were observed on February\,1 (JD\,2453402) when their FWHM and
EW were at their highest values.  However, in any case, the departures of
the jet radial velocity $V_{\rm J}$, and FWHM from their average values,
$<$\,$V_{\rm J}$\,$> =1647$\,$\pm$48\,km\,s$^{-1}$ and
$<$FWMH$>=279$\,$\pm$21\,km\,s$^{-1}$, are less than $\sim$10$\%$.  Assuming
that the jets are launched perpendicularly to the disc and orbital plane,
the jet opening angle, $\phi$, is related to the observed jet width ($\Delta
v$) and radial velocity, $\Delta v =2$\,$V_{\rm J} \sin \phi/2$,
\citep[e.g.][]{Sol1987}.  Using the average FWHM and $V_{\rm{J}}$ values, we
estimate $\phi$\,$\sim$\,$10$\degr.  The total width of the jet component is
a factor of $\sim$\,$2$ larger which may suggest a somewhat larger cone. 
However, in any case, the degree of collimation is quite high.  The jet
centre velocity, $VC_{\rm J}$, seems to follow the WD radial velocities, but
with a higher amplitude (Fig.\,\ref{F_VRCs_3}); more observations are,
however, necessary to confirm this behaviour.

The general picture for the majority of jets is that the outflow velocity
corresponds to the escape velocity at their origin \citep{Liv1998}.  Hence,
the maximum expansion velocity is set by the WD escape velocity since no
ejection can originate under its surface.  Assuming the jet expansion
perpendicular to the orbital plane, the expansion velocity is $2675\,{\rm
km\,s}^{-1}$\,$\la$\,$V_{\rm exp}$\,$=$\,$V_{\rm{J}}
\cos^{-1}i$\,$\la$\,$4816$\,km\,s$^{-1}$ for
$52$\degr\,$\la$\,$i$\,$\la$\,70\degr (Sec.~\ref{Sec-orb-sol}).  The WD
escape velocity, $v_{\rm esc}$, increases with WD mass and by the same token
with decreasing inclination ($M_{\rm
WD}$\,$\approx$\,$0.67$\,$\sin^{-3}{i}$\,M$_{\sun}$), in particular, $v_{\rm
esc}$\,$\approx$\,$5500$\,$\pm1500$\,km\,s$^{-1}$ for $M_{\rm
WD}$\,$\approx$\,$0.8$\,$\pm0.2$\,M$_\sun$ (Sec.~\ref{Sec-orb-sol}).  Hence,
the binary inclination should be close to its upper limit, $i \sim
70^{\degr}$ if the jets are launched near the WD.

\section{Epilogue}

Based on high-resolution spectroscopy combined with 16 years of optical
photometry, we have revealed the nature of the S-type symbiotic binary
St\,2-22 and derived physical parameters of the system components.  Our
double-line spectroscopic orbits indicate the orbital period $P_{\rm
{orb}}$\,$=$\,$918^{\rm d}$, the mass ratio $q$\,$=$\,$M_{\rm g}/M_{\rm
WD}$\,$=$\,$3.50$\,$\pm0.5$, and significant eccentricity
$e$\,$=$\,$0.16$\,$\pm0.07$.  The most likely orbital solution for the
component masses are $M_{\rm WD}$\,$\approx$\,$0.8$\,$\pm 0.2$\,M$_{\sun}$ and
$M_{\rm g}$\,$\approx$\,$2.8$\,$\pm0.7$\,M$_{\sun}$, whereas the orbit
inclination is $\sim$\,$70$\degr.

St\,2-22 shows outbursts with amplitudes in the range from 2--3 mag, which
seem to recur with a timescale of a dozen years.  In the quiescence, the
changes $\Delta V \approx 0\fm5$ are observed with the period P$
=277\fd8$\,$\pm1.7$.  The outburst behaviour of St\,2-22 resembles that of
the classical symbiotic binary Z\,And, and its binary parameters very
similar to those of Z\,And \citep[masses, temperatures, luminosities,
etc.;][]{MiKe1992} makes St\,2-22 almost a twin in a slightly wider orbit. 
In both systems, we observed the ejection of strongly collimated jets during
the late outburst phase.  The jet behaviour in St\,2-22, overall, resembles
that in Z\,And where gradual changes within $200$--$300$\,km\,s$^{-1}$ were
observed on a time scale of several months \citep{Sko2009}, with the
difference that the jets of St\,2-22 were undetectable in H$\beta$ (probably
due to the sensitivity of our spectra being too low); additionally, in
St\,2-22, after their first appearance in July\,2020, we observed first
deceleration and then acceleration of the jets.  The phenomenon is still
ongoing and our monitoring is continued.  We also encourage high-resolution
(R $\gtrsim 15000$) spectroscopic monitoring to document the outburst and
jet evolution well.

 \begin{figure}[t!]
   \centering
   \includegraphics[height=0.58\textwidth]{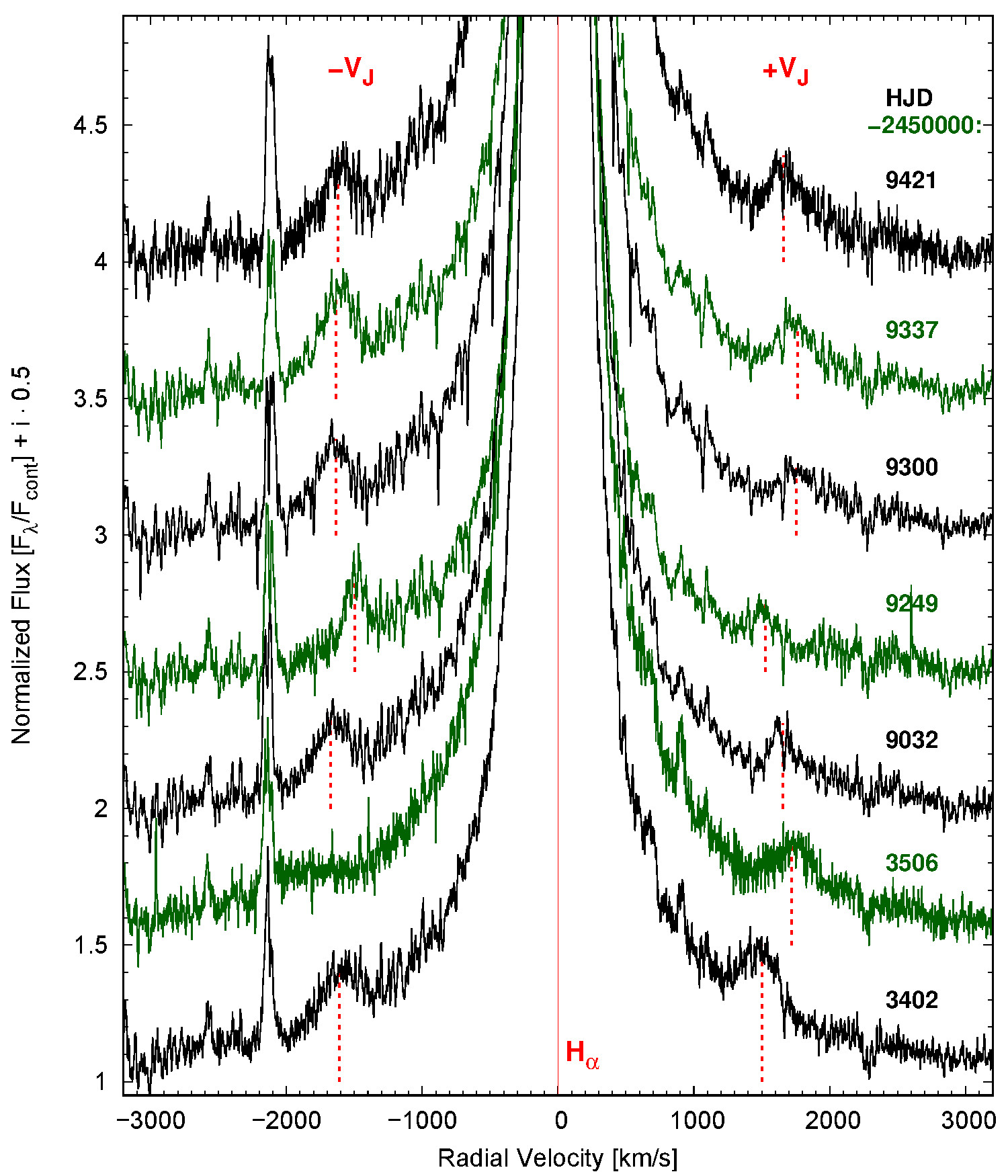}
      \caption{Evolution of H$\alpha$ line profile in the high-resolution
spectra (HRS and FEROS) of St\,2-22 with visible emission components from
collimated jets on its wings.  The measured positions of their centres
(Table\,\ref{T_jet_RV}) after correction on $V_\gamma$ are shown with red
dashed lines.
      }
         \label{F_Ha}
   \end{figure}

\begin{acknowledgements}
We dedicate this work to the memory of the wonderful man and friend, Toma
Tomov, who sadly passed away in 2019.  The present work is a continuation of
the research he initiated and CG had the honour to participate.\\ This
research has been partly financed by the Polish National Science Centre
(NCN) grants OPUS 2017/27/B/ST9/01940, MAESTRO 2015/18/A/ST9/00746, and
SONATA No.  DEC-2015/19/D/ST9/02974.  KI was supported by STFC
[ST/T000244/1].  The paper is based on spectroscopic observations made with
the Southern African Large Telescope (SALT) under programmes: 2017-1-SCI-046
and 2017-2-SCI-044 (PI: T.\,Tomov), and 2018-2-SCI-021 and 2019-1-MLT-008
(PI: C.\,Ga{\l}an).  Polish participation in SALT is funded by grant No. 
MNiSW DIR/WK/2016/07.
\end{acknowledgements}

\bibliographystyle{aa}
\bibliography{St_2-22_AaAL_20210524}


\begin{appendix} 
\section{Supplementary tables and figures}

 \begin{figure*}
   \centering
   \includegraphics[height=0.52\textwidth]{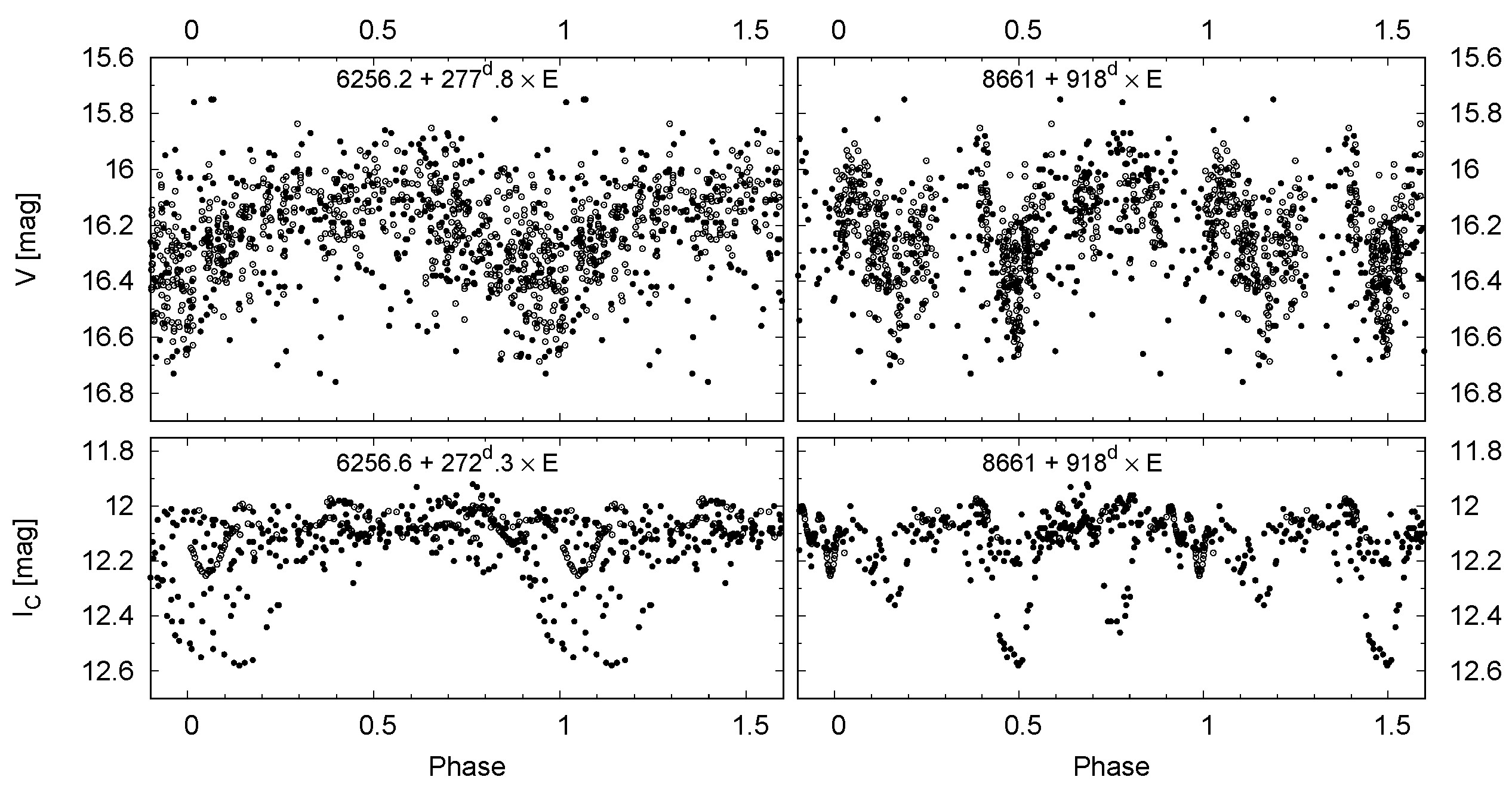}
      \caption{$V$ ({\sl Top}) and $I$ ({\sl Bottom}) light curves folded
with the shorter ({\sl Left}) and the longer ({\sl Right}) periods, respectively, with the adopted ephemerides given at the top of each plot.  
Filled and open circles correspond to the {\sl Kleinkaroo} and {\sl ASAS-SN}/{\sl OGLE} data, respectively.
      }
         \label{F_ph_LCs}
   \end{figure*}

\begin{table}[!h]
\caption{Blue cF-type absorption lines (from \ion{Ca}{i}, \ion{Sc}{ii},
\ion{Ti}{ii}, \ion{Cr}{ii}, \ion{Fe}{i}, \ion{Fe}{ii}, \ion{Y}{ii}, and
\ion{Zr}{ii}) that were identified in the high-resolution spectra collected
during outbursts and used to measure the radial velocities associated with
the hot component.}
\label{T_cF_lines}
\centering
\begin{tabular}{@{}|l@{\hskip 1mm}l@{\hskip 1mm}|l@{\hskip 1mm}l@{\hskip 1mm}|l@{\hskip 1mm}l|@{}}
\hline\hline
Element & $\lambda_{\rm {lab}}$ (Air) & Element & $\lambda_{\rm{lab}}$ (Air) & Element & $\lambda_{\rm{lab}}$ (Air) \\
        & [\AA]     &         & [\AA]      &         & [\AA]      \\
\hline
\ion{Cr}{ii} &  4242.364  & \ion{Ti}{ii} &  4394.051  & \ion{Fe}{ii} &  4534.168  \\
\ion{Sc}{ii} &  4246.822  & \ion{Ti}{ii} &  4395.031  & \ion{Fe}{ii} &  4541.524  \\
\ion{Fe}{i}  &  4271.154  & \ion{Ti}{ii} &  4395.839  & \ion{Fe}{ii} &  4549.474  \\
\ion{Fe}{i}  &  4271.761  & \ion{Ti}{ii} &  4399.765  & \ion{Ti}{ii} &  4549.622  \\
\ion{Fe}{ii} &  4273.320  & \ion{Sc}{ii} &  4400.389  & \ion{Fe}{ii} &  4555.893  \\
\ion{Fe}{ii} &  4273.325  & \ion{Fe}{i}  &  4404.750  & \ion{Cr}{ii} &  4558.650  \\
\ion{Cr}{ii} &  4275.567  & \ion{Ti}{ii} &  4407.672  & \ion{Ti}{ii} &  4563.757  \\
\ion{Fe}{i}  &  4282.403  & \ion{Fe}{i}  &  4407.709  & \ion{Ti}{ii} &  4571.971  \\
\ion{Ti}{ii} &  4287.873  & \ion{Ti}{ii} &  4409.235  & \ion{Fe}{ii} &  4576.340  \\
\ion{Fe}{ii} &  4296.572  & \ion{Ti}{ii} &  4409.516  & \ion{Fe}{ii} &  4582.835  \\
\ion{Ti}{ii} &  4300.042  & \ion{Ti}{ii} &  4411.072  & \ion{Fe}{ii} &  4583.837  \\
\ion{Ti}{ii} &  4301.914  & \ion{Ti}{ii} &  4411.929  & \ion{Cr}{ii} &  4588.199  \\
\ion{Sc}{ii} &  4305.714  & \ion{Ti}{ii} &  4418.330  & \ion{Ti}{ii} &  4589.958  \\
\ion{Ti}{ii} &  4307.863  & \ion{Ti}{ii} &  4421.938  & \ion{Cr}{ii} &  4592.049  \\
\ion{Fe}{i}  &  4307.902  & \ion{Fe}{i}  &  4422.568  & \ion{Fe}{ii} &  4629.339  \\
\ion{Ti}{ii} &  4312.860  & \ion{Y}{ii}  &  4422.585  & \ion{Cr}{ii} &  4634.070  \\
\ion{Sc}{ii} &  4314.083  & \ion{Ca}{i}  &  4434.957  & \ion{Ti}{ii} &  4763.883  \\
\ion{Fe}{ii} &  4314.310  & \ion{Ca}{i}  &  4435.679  & \ion{Ti}{ii} &  4764.526  \\
\ion{Ti}{ii} &  4314.971  & \ion{Ti}{ii} &  4441.729  & \ion{Ti}{ii} &  4779.985  \\
\ion{Fe}{i}  &  4315.085  & \ion{Fe}{i}  &  4442.339  & \ion{Ti}{ii} &  4798.521  \\
\ion{Sc}{ii} &  4320.732  & \ion{Ti}{ii} &  4443.794  & \ion{Ti}{ii} &  4805.085  \\
\ion{Ti}{ii} &  4320.950  & \ion{Ti}{ii} &  4444.555  & \ion{Cr}{ii} &  4812.337  \\
\ion{Sc}{ii} &  4324.996  & \ion{Ti}{ii} &  4450.482  & \ion{Cr}{ii} &  4824.127  \\
\ion{Fe}{i}  &  4325.762  & \ion{Ca}{i}  &  4454.779  & \ion{Cr}{ii} &  4836.229  \\
\ion{Ti}{ii} &  4330.237  & \ion{Fe}{i}  &  4461.653  & \ion{Cr}{ii} &  4848.235  \\
\ion{Ti}{ii} &  4330.240  & \ion{Ti}{ii} &  4464.449  & \ion{Ti}{ii} &  4874.010  \\
\ion{Ti}{ii} &  4330.695  & \ion{Ti}{ii} &  4468.507  & \ion{Cr}{ii} &  4876.399  \\
\ion{Ti}{ii} &  4344.281  & \ion{Fe}{ii} &  4489.183  & \ion{Cr}{ii} &  4876.410  \\
\ion{Sc}{ii} &  4354.598  & \ion{Fe}{ii} &  4491.405  & \ion{Y}{ii}  &  4900.119  \\
\ion{Ti}{ii} &  4367.652  & \ion{Ti}{ii} &  4501.270  & \ion{Ti}{ii} &  4911.195  \\
\ion{Sc}{ii} &  4374.457  & \ion{Fe}{ii} &  4508.288  & \ion{Fe}{ii} &  4923.927  \\
\ion{Ti}{ii} &  4374.816  & \ion{Fe}{ii} &  4515.339  & \ion{Fe}{ii} &  5018.440  \\
\ion{Y}{ii}  &  4374.933  & \ion{Ti}{ii} &  4518.327  & \ion{Sc}{ii} &  5031.021  \\
\ion{Zr}{ii} &  4379.742  & \ion{Fe}{ii} &  4520.224  & \ion{Ti}{ii} &  5072.287  \\
\ion{Sc}{ii} &  4384.814  & \ion{Fe}{ii} &  4522.634  & \ion{Ti}{ii} &  5185.902  \\
\ion{Fe}{ii} &  4385.387  & \ion{Ti}{ii} &  4529.480  & \ion{Ti}{ii} &  5188.687  \\
\ion{Ti}{ii} &  4391.025  & \ion{Ti}{ii} &  4533.960  &              &            \\
\hline
\end{tabular}
\end{table}

\begin{table}[!h]
\caption{Emission lines identified in the HRS/SALT blue- and red-arm spectra
of St\,2-22.}
\label{T_emission_lines_II}
\centering
\begin{tabular}{|l@{\hskip 1mm}|l@{\hskip 1mm}|l@{\hskip 1mm}|}
\hline\hline
Line                                                                 & Line                                         & Line                                        \\
\hline
H$\delta$ ($\lambda$\,4101.7\,\AA)                                   & \ion{Fe}{ii} ($\lambda$\,5040.8\,\AA)        & \ion{Fe}{ii} ($\lambda$\,6370.3\,\AA)       \\
\ion{Fe}{ii} ($\lambda$\,4233.2\,\AA)                                & \ion{He}{i} ($\lambda$\,5047.7\,\AA)         & \ion{Fe}{ii} ($\lambda$\,6383.7\,\AA)       \\
H$\gamma$ ($\lambda$\,4341.3\,\AA)                                   & \ion{Fe}{ii} ($\lambda$\,5057.0\,\AA)        & \ion{Fe}{ii} ($\lambda$\,6385.5\,\AA)       \\
$[$\ion{O}{iii}$]$ ($\lambda$\,4363.8\,\AA)                          & \ion{Fe}{ii} ($\lambda$\,5100.7\,\AA)        & \ion{Fe}{ii} ($\lambda$\,6416.9\,\AA)       \\
\ion{He}{i} ($\lambda$\,4387.9\,\AA)                                 & \ion{Fe}{ii} ($\lambda$\,5132.6\,\AA)        & \ion{Fe}{ii} ($\lambda$\,6432.7\,\AA)       \\
\ion{He}{i} ($\lambda$\,4471.5\,\AA)                                 & \ion{Fe}{ii} ($\lambda$\,5145.8\,\AA)        & \ion{Fe}{ii} ($\lambda$\,6456.4\,\AA)       \\
\ion{Mg}{ii} ($\lambda$\,4481.2\,\AA)                                & $[$\ion{Fe}{vii}$]$ ($\lambda$\,5158.9\,\AA) & \ion{Fe}{ii} ($\lambda$\,6516.1\,\AA)       \\
\ion{Fe}{ii} ($\lambda$\,4491.4\,\AA)                                & \ion{Fe}{ii} ($\lambda$\,5169.0\,\AA)        & H$\alpha$ ($\lambda$\,6562.8\,\AA)          \\
\ion{Fe}{ii} ($\lambda$\,4508.3\,\AA)                                & \ion{Fe}{ii} ($\lambda$\,5197.6\,\AA)        & \ion{He}{i} ($\lambda$\,6678.2\,\AA)        \\
\ion{Fe}{ii} ($\lambda$\,4515.3\,\AA)                                & \ion{Fe}{ii} ($\lambda$\,5234.6\,\AA)        & \ion{O}{vi} ($\lambda$\,6832\,\AA)          \\
\ion{Fe}{ii} ($\lambda$\,4522.6\,\AA)                                & \ion{Fe}{ii} ($\lambda$\,5254.4\,\AA)        & \ion{He}{i} ($\lambda$\,7065.2\,\AA)        \\
\ion{He}{ii} ($\lambda$\,4542\,\AA)                                  & $[$\ion{Fe}{ii}$]$ ($\lambda$\,5262.3\,\AA)  & $[$\ion{Fe}{ii}$]$ ($\lambda$\,7155.2\,\AA) \\
\ion{Fe}{ii} ($\lambda$\,4555.9\,\AA)                                & \ion{Fe}{ii} ($\lambda$\,5264.8\,\AA)        & \ion{He}{i} ($\lambda$\,7281.3\,\AA)        \\
\ion{Mg}{i}$]$ ($\lambda$\,4571.1\,\AA)                              & \ion{Fe}{ii} ($\lambda$\,5276.0\,\AA)        & \ion{Fe}{ii} ($\lambda$\,7462.4\,\AA)       \\
\ion{Fe}{ii} ($\lambda$\,4576.3\,\AA)                                & \ion{Fe}{ii} ($\lambda$\,5284.1\,\AA)        & \ion{Fe}{ii} ($\lambda$\,7516.2\,\AA)       \\
\ion{Fe}{ii} ($\lambda$\,4583.8\,\AA)                                & \ion{Fe}{ii} ($\lambda$\,5316.6\,\AA)        & \ion{Fe}{ii} ($\lambda$\,7711.4\,\AA)       \\
\ion{Fe}{ii} ($\lambda$\,4620.5\,\AA)                                & \ion{Fe}{ii} ($\lambda$\,5363.0\,\AA)        & \ion{Fe}{ii} ($\lambda$\,7866.6\,\AA)       \\
\ion{Fe}{ii} ($\lambda$\,4629.3\,\AA)                                & \ion{He}{ii} ($\lambda$\,5411.5\,\AA)        & \ion{Fe}{ii} ($\lambda$\,7877.0\,\AA)       \\
\ion{N}{iii} ($\lambda$\,4630.6\,\AA)\tablefootmark{$\blacklozenge$} & \ion{Fe}{ii} ($\lambda$\,5414.9\,\AA)        & \ion{Mg}{ii} ($\lambda$\,7896.4\,\AA)       \\
\ion{N}{iii} ($\lambda$\,4642.0\,\AA)\tablefootmark{$\blacklozenge$} & \ion{Fe}{ii} ($\lambda$\,5425.3\,\AA)        & \ion{H}{i} ($\lambda$\,8333.8\,\AA)         \\
\ion{C}{iii} ($\lambda$\,4647.4\,\AA)\tablefootmark{$\blacklozenge$} & \ion{Fe}{ii} ($\lambda$\,5534.9\,\AA)        & \ion{H}{i} ($\lambda$\,8345.5\,\AA)         \\
\ion{C}{iii} ($\lambda$\,4652.0\,\AA)\tablefootmark{$\blacklozenge$} & $[$\ion{O}{i}$]$ ($\lambda$\,5577.3\,\AA)    & \ion{H}{i} ($\lambda$\,8359.0\,\AA)         \\
\ion{Fe}{ii} ($\lambda$\,4640.8\,\AA)                                & $[$\ion{Ca}{vii}$]$ ($\lambda$\,5618.8\,\AA) & \ion{H}{i} ($\lambda$\,8374.5\,\AA)         \\
$[$\ion{Fe}{iii}$]$ ($\lambda$\,4658.1\,\AA)                         & $[$\ion{Fe}{vii}$]$ ($\lambda$\,5720.7\,\AA) & \ion{H}{i} ($\lambda$\,8392.4\,\AA)         \\
\ion{Fe}{ii} ($\lambda$\,4666.7\,\AA)                                & \ion{He}{i} ($\lambda$\,5875.6\,\AA)         & \ion{H}{i} ($\lambda$\,8413.3\,\AA)         \\
\ion{Fe}{ii} ($\lambda$\,4670.2\,\AA)                                & \ion{Si}{ii} ($\lambda$\,5978.9\,\AA)        & \ion{H}{i} ($\lambda$\,8438.0\,\AA)         \\
\ion{He}{ii} ($\lambda$\,4685.7\,\AA)                                & \ion{Fe}{ii} ($\lambda$\,5991.4\,\AA)        & \ion{O}{i} ($\lambda$\,8446.4\,\AA)         \\
\ion{He}{i} ($\lambda$\,4713.2\,\AA)                                 & \ion{Fe}{ii} ($\lambda$\,6084.1\,\AA)        & \ion{H}{i} ($\lambda$\,8467.3\,\AA)         \\
\ion{Fe}{ii} ($\lambda$\,4731.4\,\AA)                                & $[$\ion{Fe}{vii}$]$ ($\lambda$\,6087.0\,\AA) & \ion{Ca}{ii} ($\lambda$\,8498.0\,\AA)       \\
H$\beta$ ($\lambda$\,4862.9\,\AA)                                    & \ion{Fe}{ii} ($\lambda$\,6147.7\,\AA)        & \ion{H}{i} ($\lambda$\,8502.5\,\AA)         \\
\ion{He}{i} ($\lambda$\,4921.9\,\AA)                                 & \ion{Fe}{ii} ($\lambda$\,6238.4\,\AA)        & \ion{Ca}{ii} ($\lambda$\,8542.1\,\AA)       \\
\ion{Fe}{ii} ($\lambda$\,4924.3\,\AA)                                & \ion{Fe}{ii} ($\lambda$\,6247.6\,\AA)        & \ion{H}{i} ($\lambda$\,8545.4\,\AA)         \\
\ion{Fe}{ii} ($\lambda$\,4959.6\,\AA)                                & $[$\ion{O}{i}$]$ ($\lambda$\,6300.3\,\AA)    & \ion{H}{i} ($\lambda$\,8598.4\,\AA)         \\
\ion{Fe}{ii} ($\lambda$\,4993.4\,\AA)                                & \ion{He}{ii} ($\lambda$\,6311\,\AA)          & \ion{Ca}{ii} ($\lambda$\,8662.1\,\AA)       \\
$[$\ion{O}{iii}$]$ ($\lambda$\,5006.8\,\AA)                          & \ion{Fe}{ii} ($\lambda$\,6317.4\,\AA)        & \ion{H}{i} ($\lambda$\,8665.0\,\AA)         \\
\ion{He}{i} ($\lambda$\,5015.7\,\AA)                                 & \ion{Fe}{ii} ($\lambda$\,6347.6\,\AA)        & \ion{H}{i} ($\lambda$\,8750.5\,\AA)         \\
\ion{Fe}{ii} ($\lambda$\,5018.4\,\AA)                                & $[$\ion{O}{i}$]$ ($\lambda$\,6363.8\,\AA)    &                                             \\
\hline
\end{tabular}
\tablefoot{\tablefoottext{$\blacklozenge$}{\tiny {These are bands composed of several lines. The initial and the ending wavelengths are only given.}}\\
          }
\end{table}

\begin{figure}[!h]
   \centering
   \includegraphics[height=0.34\textwidth]{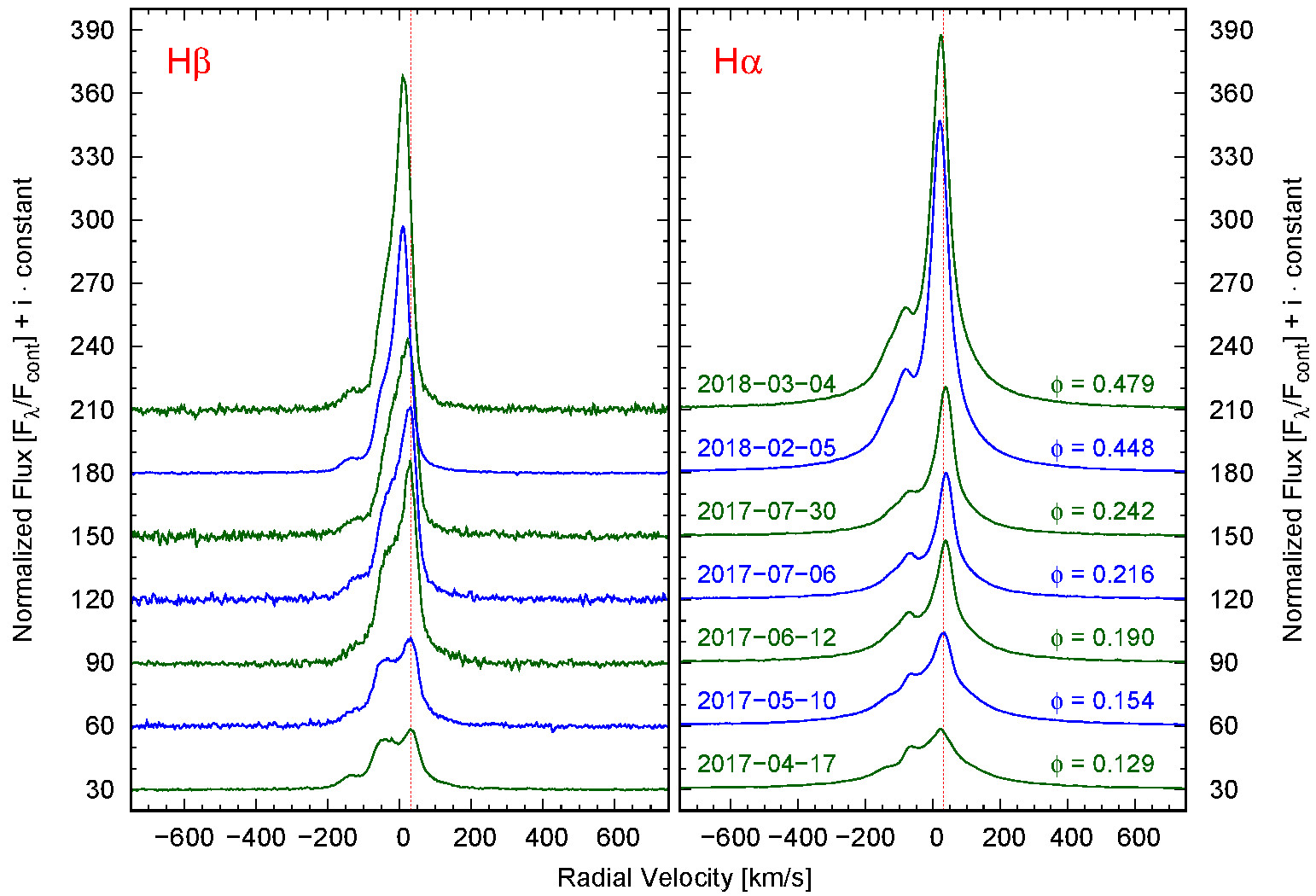}
      \caption{Evolution of the H$\beta$ ({\sl  {Left}}) and H$\alpha$ ({\sl
{Right}}) line profiles in the HRS/SALT spectra (Low-resolution mode: R
$\sim 14000$) collected during quiescence.  The red dashed vertical lines
mark the systemic velocity.}
         \label{F_HaHb_q}
   \end{figure}

 \begin{figure}[!h]
   \centering
   \includegraphics[height=0.56\textwidth]{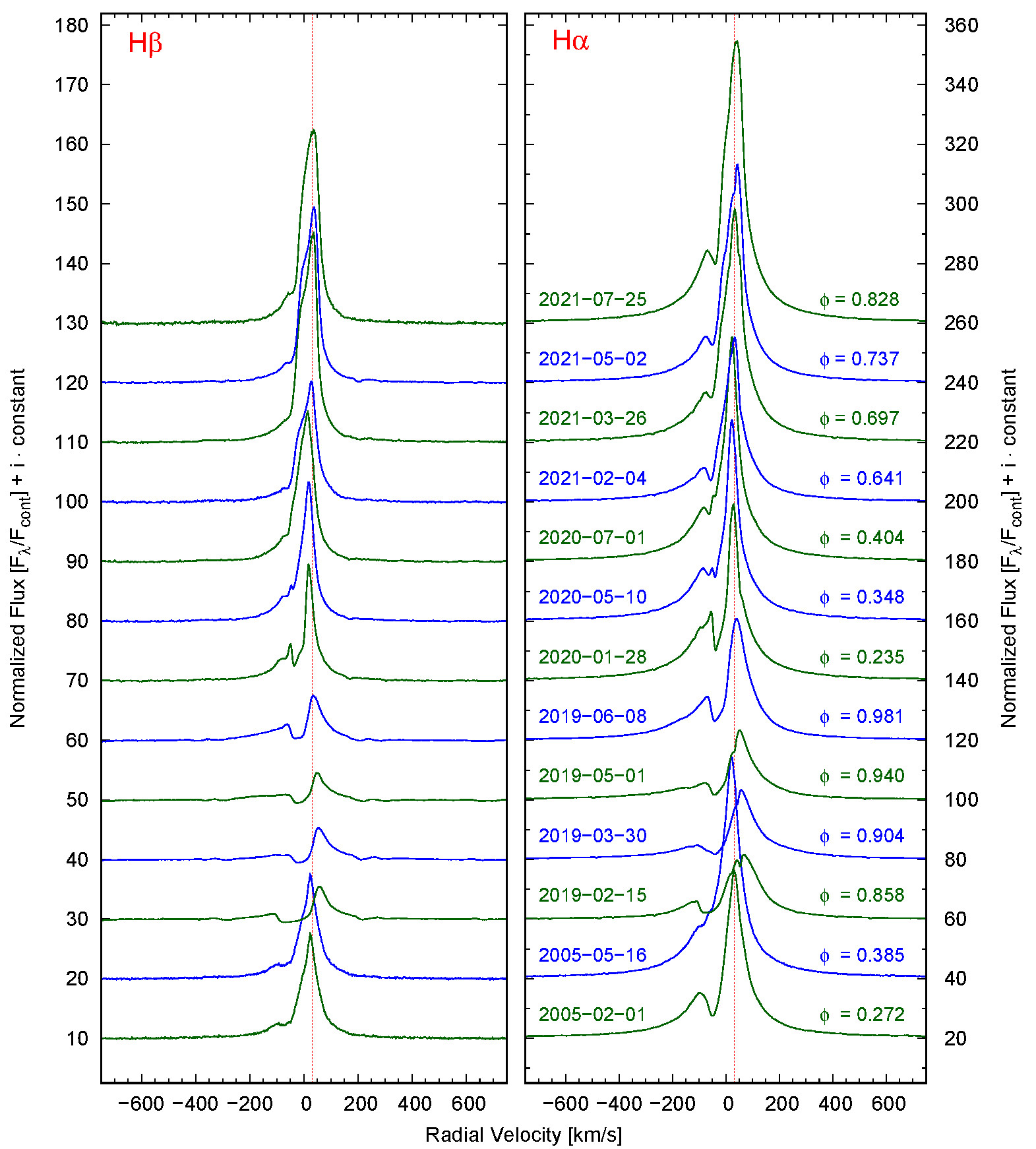}
      \caption{Evolution of the H$\beta$ ({\sl {Left}}) and H$\alpha$ ({\sl
{Right}}) line profiles during the last outburst in the HRS/SALT spectra
(medium-resolution mode: R $\sim 40000$) and two FEROS spectra taken during
the previous outburst in 2005.  The red dashed vertical lines mark the
systemic velocity.}
         \label{F_HaHb_o}
   \end{figure}

 \begin{figure*}[!h]
   \centering
   \includegraphics[height=0.65\textwidth]{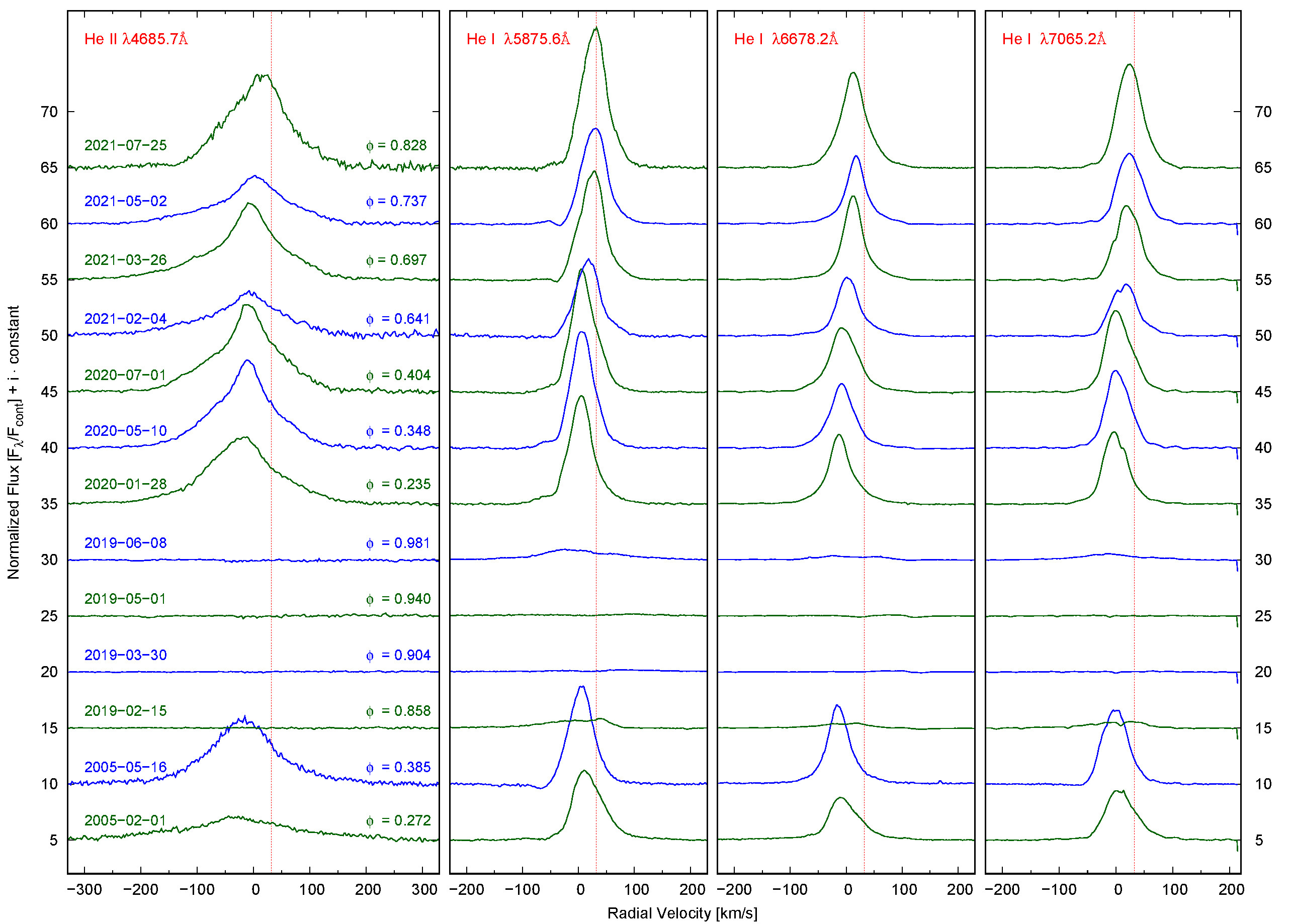}
      \caption{Evolution of the \ion{He}{ii} ($\lambda 4685.680$\,\AA) and
\ion{He}{i} ($\lambda 5875.640$\,\AA, $\lambda 6678.151$\,\AA, and $\lambda
7065.190$\,\AA) line profiles during the present outburst in the HRS/SALT
spectra (medium-resolution mode: R $\sim 40000$) and two FEROS spectra taken
during the 2005 outburst.  The red dashed vertical lines mark
the systemic velocity.}
         \label{F_He_II_I_o}
   \end{figure*}

 \begin{figure}[!h]
   \centering
    \includegraphics[height=0.85\textwidth]{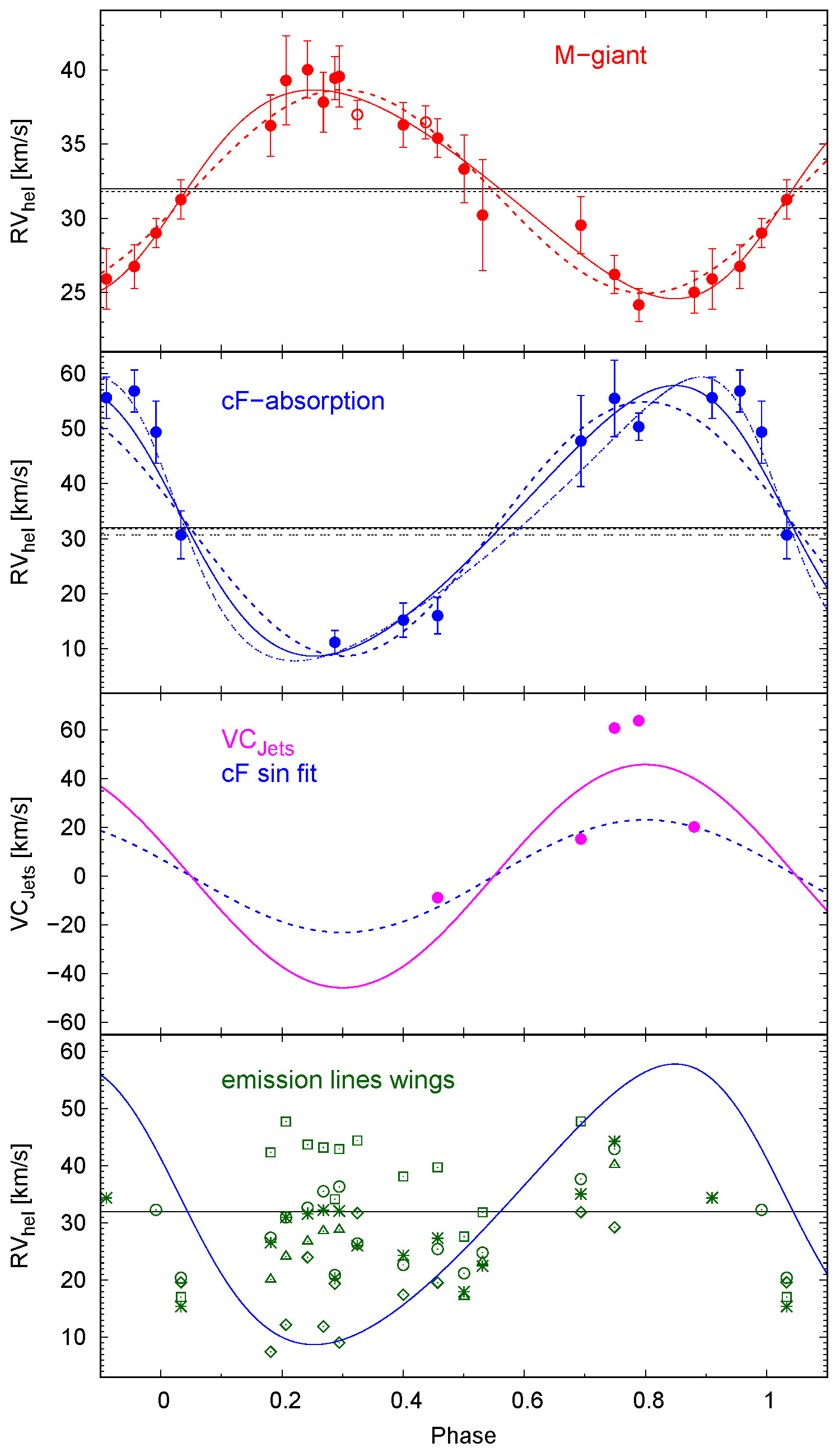}
      \caption{Radial velocity curves of the red gaint ({\sl {top}}) and hot
               ({\sl {middle}}) components folded with period P$_{\rm {sp}}
               = 918^{\rm d}$.  SALT/HRS and {\sl {FEROS}} data are shown
               with filled and open circles, respectively.  Solid lines show
               synthetic radial velocity curves for the case of the
               eccentric orbit and dashed lines in the case of a circular
               orbit.  The horizontal lines represent the systemic
               velocities for the eccentric (solid) and circular (dashed)
               cases.  The lower {\sl {middle}} panel shows the jet centre
               velocity $VC_{\rm J}$.  The {\sl {bottom}} panel presents the
               radial velocities of the emission line wings: H$\alpha$
               (circles), H$\beta$ (diamonds), \ion{He}{i} $\lambda
               5875.6$\,\AA\  (asterisks), \ion{He}{ii} $\lambda
               5411.5$\,\AA\, (triangles) and the mean value from
               \ion{He}{i} $\lambda 5875.6$\,\AA, and $\lambda
               7065.2$\,\AA\, lines (squares), respectively, compared to the
               radial velocity curve of the hot component (blue line).  The
               method described by \citet {Sha1983} and references therein
               were used to measure the velocities of emission lines wings.
}
         \label{F_VRCs_3}
   \end{figure}

\end{appendix}

\end{document}